\documentclass[aps,pre,preprint,nofootinbib]{revtex4}

\usepackage{amsmath,amssymb,amsfonts,graphicx,framed}
\usepackage{array}

\begin{document}

\title{Constructing Separable Non-$2\pi$-Periodic Solutions to the Navier-Lam\'{e} Equation in Cylindrical Coordinates Using the Buchwald Representation: Theory and Applications}

\author{Jamal Sakhr}
\author{Blaine A. Chronik}
\affiliation{Department of Physics and Astronomy, The University of Western Ontario, London, Ontario N6A 3K7 Canada}

\date{\today}

\begin{abstract}
In a previous paper [Adv. Appl. Math. Mech. 10 (2018), pp. 1025-1056], we used the Buchwald representation to construct several families of separable cylindrical solutions to the Navier-Lam\'{e} equation; these solutions had the property of being $2\pi$-periodic in the circumferential coordinate. In this paper, we extend the analysis and obtain the complementary set of separable solutions whose circumferential parts are elementary $2\pi$-aperiodic functions. Collectively, we construct eighteen distinct families of separable solutions; in each case, the circumferential part of the solution is one of three elementary $2\pi$-aperiodic functions. These solutions are useful for solving a wide variety of dynamical problems that involve cylindrical geometries and for which $2\pi$-periodicity in the angular coordinate is incompatible with the given boundary conditions. As illustrative examples, we show how the obtained solutions can be used to solve certain forced-vibration problems involving open cylindrical shells and open solid cylinders where (by virtue of the boundary conditions) $2\pi$-periodicity in the angular coordinate is inappropriate. As an addendum to our prior work, we also include an illustrative example of a certain type of asymmetric problem that can be solved using the particular $2\pi$-periodic subsolutions that ensue when there is no explicit dependence on the circumferential coordinate. 
\end{abstract}

\maketitle

\section{Introduction}

In linear elastodynamics \cite{Elastobuch}, the displacement of a homogeneous, isotropic, and linearly-elastic solid is governed by the Navier-Lam\'{e} (NL) equation. Over the last century, many specialized methods have been developed for obtaining solutions to the NL equation. One of the most popular and effective methods is the method of potentials \cite{Gurtin,Miky,Ruskie}, whereby the displacement vector is represented as a specific combination of one or more scalar and/or vector functions (called potentials). The scalar components of the NL equation are a set of coupled PDEs that cannot generally be solved in closed form. Representations in terms of displacement potentials may, once substituted into the NL equation, yield a system of PDEs that are uncoupled, less coupled, or at least simpler than the original component equations. Many different representations exist including the classical Helmholtz-Lam\'{e} and Papkovich-Neuber representations. 
In applications, the analytical utility of a representation often depends on the choice of the working coordinate system. One representation that has proven to be analytically effective in anisotropic problems with cylindrical symmetry is the so-called Buchwald representation (see Refs.~\cite{AR98,AR00,AR01} and references therein). The Buchwald representation is also applicable to problems involving isotropic media \cite{thesis2,ArXpap}, but its application to such problems is relatively rare \cite{thesis1,Indguys1,Indguys2}. 

The Buchwald representation involves three scalar potential functions (see Eq.~(\ref{Buchy}) of Sec.~\ref{EOMsALL}). It can be shown by various means (see, for example, Refs.~\cite{thesis1,usB1}) that the Buchwald representation reduces the original scalar components of the NL equation (which are a set of three coupled PDEs) to a set of two coupled PDEs involving two of the potentials and one separate decoupled PDE involving the remaining potential. By assuming separable product solutions for the three potentials and imposing certain conditions on their axial and temporal parts, it was shown in Ref.~\cite{usB1} that the coupled subsystem reduces to a homogeneous linear system that can be solved by elimination. The elimination procedure generates two independent fourth-order linear PDEs with constant coefficients, which can be subsequently solved using linear combinations of solutions to the two-dimensional Helmholtz equation in polar coordinates. A solution to the independent decoupled PDE can be obtained using separation of variables. In Ref.~\cite{usB1} (henceforth referred to as SC1), we specifically constructed particular solutions (to the PDEs governing the Buchwald potentials) possessing $2\pi$-periodic angular parts. In the present paper, we consider the complementary class of separable solutions whose angular parts are either aperiodic or periodic but not $2\pi$-periodic. Although more involved, all three Buchwald potentials can again be completely determined (under the stipulated conditions), and following the approach of SC1 we can then construct eighteen distinct families of parametric solutions to the NL equation. The angular parts of these solutions (to the NL equation) are of course \emph{not} $2\pi$-periodic. One new complication that arises is that some solutions lacking $2\pi$-periodicity in their angular parts are complex-valued, and in those cases, we discuss how to extract acceptable real-valued solutions. 

To our knowledge, the solutions obtained here constitute the first comprehensive set of $2\pi$-aperiodic cylindrical solutions to the NL equation. 
Independent of the fact that they were derived using Buchwald potentials, these exact closed-form families of solutions are fundamentally important and useful in their own right. They can be directly applied to a fundamental set of linear-elastic boundary-value-problems in cylindrical coordinates where $2\pi$-periodicity in the angular coordinate is incompatible with the given boundary conditions. In the second half of the paper, we provide three illustrative examples of such problems and detail how the obtained parametric solutions can be used to solve them. As an addendum to SC1, we also include an example that illustrates the use of a special set of $2\pi$-periodic subsolutions that results when there is no explicit functional dependence on the circumferential coordinate. 

\section{Equations of Motion}\label{EOMsALL} 

The NL equation can be written in vector form as \cite{Elastobuch,Ruskie}
\begin{subequations}\label{NLE} 
\begin{equation}\label{NLE1}
\mu\nabla^2\mathbf{u}+(\lambda+\mu)\nabla(\nabla\cdot\mathbf{u})+\mathbf{b}=\rho{\partial^2\mathbf{u}\over\partial t^2},
\end{equation}
or (using the well-known identity for the vector Laplacian\footnote{$\displaystyle \nabla^2\mathbf{u}=\nabla(\nabla\cdot\mathbf{u})-\nabla\times(\nabla\times\mathbf{u})$}) as 
\begin{equation}\label{NLE2}
(\lambda+2\mu)\nabla(\nabla\cdot\mathbf{u})-\mu\nabla\times(\nabla\times\mathbf{u})+\mathbf{b}=\rho{\partial^2\mathbf{u}\over\partial t^2},
\end{equation}
\end{subequations} 
where $\mathbf{u}$ is the displacement vector, $\lambda$ and $\mu$ are the Lam\'{e} constants from the classical linear theory of elasticity, $\mathbf{b}$ is the local body force, and $\rho$ is the (constant) density of the material. In this paper,  we consider the situation in which there are only surface forces acting on the solid and no body forces (i.e., $\mathbf{b}=0$). In the circular cylindrical coordinate system, the displacement $\mathbf{u}\equiv\mathbf{u}(r,\theta,z,t)$ depends on the spatial coordinates $(r,\theta,z)$, which denote the radial, circumferential, and longitudinal coordinates, respectively, and on the time $t$. 

The (standard) Buchwald representation of the displacement field involves three scalar potentials $\Phi(r,\theta,z,t)$, $\Psi(r,\theta,z,t)$, $\chi(r,\theta,z,t)$, and is written as follows \cite{AR98}: 
\begin{equation}\label{Buchy}
\mathbf{u}=\nabla\Phi+\nabla\times(\chi\mathbf{\hat{z}})+\left({\partial\Psi\over\partial z}-{\partial\Phi\over\partial z}\right)\mathbf{\hat{z}},
\end{equation}
where $\mathbf{\hat{z}}$ is the unit basis vector in the $z$-direction. 
It can be shown by various means \cite{thesis1,usB1} that representation (\ref{Buchy}) satisfies Eq.~(\ref{NLE}) identically (assuming $\mathbf{b}=0$) provided the Buchwald potentials $\Phi,\Psi$, and $\chi$ are solutions of the coupled PDE system: 
\begin{subequations}\label{systemeSFIN} 
\begin{equation}\label{systemeSFIN1}
(\lambda+2\mu)\nabla^2\Phi+(\lambda+\mu){\partial^2\Psi\over\partial z^2}-(\lambda+\mu){\partial^2\Phi\over\partial z^2}-\rho{\partial^2\Phi\over\partial t^2}=0,
\end{equation}
\begin{equation}\label{systemeSFIN2}
(\lambda+\mu)\left[\nabla^2\Phi-{\partial^2\Phi\over\partial z^2}\right] + \mu\nabla^2\Psi + (\lambda+\mu){\partial^2\Psi\over\partial z^2} - \rho{\partial^2\Psi\over\partial t^2}=0. 
\end{equation}
\begin{equation}\label{systemeSFIN3}
\mu\nabla^2\chi-\rho{\partial^2\chi\over\partial t^2}=0.
\end{equation}
\end{subequations} 
 
In SC1 \cite{usB1}, we obtained, under certain conditions, particular separable solutions to subsystem (\ref{systemeSFIN1},\ref{systemeSFIN2}) that are $2\pi$-periodic in the circumferential coordinate $\theta$. In the present paper, we seek to obtain the complementary set of separable solutions to subsystem (\ref{systemeSFIN1},\ref{systemeSFIN2}) whose circumferential functions are $2\pi$-aperiodic. By construction, the general method of solution will be guided by and closely follow the method developed in SC1. 

\section{Particular Separable Solutions}\label{PartSolns} 

Assume that the Buchwald potentials $\Phi$ and $\Psi$ have the separable form:
\begin{subequations}\label{solform} 
\begin{equation}\label{solform1}
\Phi(r,\theta,z,t)=\phi_\perp(r,\theta)\phi_z(z)\phi_t(t),
\end{equation}
\begin{equation}\label{solform2}
\Psi(r,\theta,z,t)=\psi_\perp(r,\theta)\psi_z(z)\psi_t(t),
\end{equation}
\end{subequations}
which together with certain conditions on the functions $\displaystyle \{\phi_z(z),\phi_t(t),\psi_z(z),\psi_t(t)\}$ (to be specified below) define the class of separable solutions to subsystem (\ref{systemeSFIN1},\ref{systemeSFIN2}) that we seek to find in this paper. Using the notation
\begin{equation}\label{2DLap}
\nabla_\perp^2\equiv\nabla^2-{\partial^2\over\partial z^2}={\partial^2\over\partial r^2} + {1\over r}{\partial\over\partial r} + {1\over r^2}{\partial^2\over\partial \theta^2}
\end{equation}
to denote the two-dimensional Laplacian operator in polar coordinates, subsystem (\ref{systemeSFIN1},\ref{systemeSFIN2}) can then be cast in the form:
\begin{subequations}\label{systemeF} 
\begin{equation}\label{systemeF1}
(\lambda+2\mu)\nabla_\perp^2\Phi+\mu{\partial^2\Phi\over\partial z^2} + (\lambda+\mu){\partial^2\Psi\over\partial z^2} - \rho{\partial^2\Phi\over\partial t^2}=0,
\end{equation}
\begin{equation}\label{systemeF2}
(\lambda+\mu)\nabla_\perp^2\Phi+\mu\nabla_\perp^2\Psi+(\lambda+2\mu){\partial^2\Psi\over\partial z^2}-\rho{\partial^2\Psi\over\partial t^2}=0.
\end{equation}
\end{subequations}
Substituting (\ref{solform}) into system (\ref{systemeF}) leads to the system
\begin{subequations}\label{systemesub} 
\begin{equation}\label{systemesub1}
\left[(\lambda+2\mu)\phi_z\phi_t\nabla_\perp^2+\mu\phi_z''\phi_t - \rho\phi_z\phi_t''\right]\phi_\perp + \left[(\lambda+\mu)\psi_z''\psi_t\right]\psi_\perp=0,
\end{equation}
\begin{equation}\label{systemesub2}
\left[(\lambda+\mu)\phi_z\phi_t\nabla_\perp^2\right]\phi_\perp+\left[\mu\psi_z\psi_t\nabla_\perp^2+(\lambda+2\mu)\psi_z''\psi_t-\rho\psi_z\psi_t''\right]\psi_\perp=0.
\end{equation}
\end{subequations}
A special class of solutions can be obtained by assuming common functional dependences in the axial and temporal parts of both potentials, that is, 
\begin{subequations}\label{assump} 
\begin{equation}\label{assump3}
\phi_z(z)=\psi_z(z),
\end{equation}
\begin{equation}\label{assump6}
\phi_t(t)=\psi_t(t),
\end{equation}
and by further assuming that the derivatives of the above functions are such that
\begin{equation}\label{assump1}
\phi_z''(z)=\kappa\phi_z(z),
\end{equation}
\begin{equation}\label{assump2}
\psi_z''(z)=\kappa\psi_z(z),
\end{equation}
\begin{equation}\label{assump4}
\phi_t''(t)=\tau\phi_t(t),
\end{equation}
\begin{equation}\label{assump5}
\psi_t''(t)=\tau\psi_t(t),
\end{equation}
\end{subequations}
where $\displaystyle \kappa\in\mathbb{R}$ and $\displaystyle \tau\in\mathbb{R}\backslash\{0\}$ are arbitrary constants that we shall refer to as the axial and time constants, respectively. Ultimately, they will appear as parameters in the final solution. With these assumptions system (\ref{systemesub}) reduces to
\begin{subequations}\label{systemefin} 
\begin{equation}\label{systemefin1}
\left[(\lambda+2\mu)\nabla_\perp^2+\mu\kappa - \rho\tau\right]\phi_\perp + \left[(\lambda+\mu)\kappa\right]\psi_\perp=0,
\end{equation}
\begin{equation}\label{systemefin2}
\left[(\lambda+\mu)\nabla_\perp^2\right]\phi_\perp+\left[\mu\nabla_\perp^2+(\lambda+2\mu)\kappa-\rho\tau\right]\psi_\perp=0, 
\end{equation}
\end{subequations}
which has the form of a homogeneous linear system:
\begin{subequations}\label{systemefinL} 
\begin{equation}\label{systemefinL1}
L_1\phi_\perp + L_2\psi_\perp=0,
\end{equation}
\begin{equation}\label{systemefinL2}
L_3\phi_\perp+L_4\psi_\perp=0, 
\end{equation}
where the linear operators $\{L_1,L_2,L_3,L_4\}$ 
\begin{eqnarray}
L_1&=&(\lambda+2\mu)\nabla_\perp^2+\mu\kappa - \rho\tau, \\
L_2&=&(\lambda+\mu)\kappa, \\
L_3&=&(\lambda+\mu)\nabla_\perp^2, \\
L_4&=&\mu\nabla_\perp^2+(\lambda+2\mu)\kappa-\rho\tau,
\end{eqnarray}
\end{subequations}
are each first-order polynomial differential operators in $\nabla_\perp^2$ with constant coefficients. Two distinct cases can be distinguished at this point depending on whether $\kappa\neq0$ or $\kappa=0$. The former is the general case while the latter is a special case for which the linear system (\ref{systemefinL}) decouples. We shall consider these two cases separately. 

\subsection{General Case: $\displaystyle\kappa\neq0$}\label{solnkapneq0}

Under conditions (\ref{assump}) and assuming $\kappa\neq0$, we can solve for $\phi_\perp(r,\theta)$ and $\psi_\perp(r,\theta)$ using the method of elimination. It can be shown that $\phi_\perp(r,\theta)$ and $\psi_\perp(r,\theta)$ independently satisfy the decoupled equations: 
\begin{subequations}\label{detty} 
\begin{eqnarray}\label{detty1}
(L_1L_4-L_2L_3)\phi_\perp=0,
\end{eqnarray}
\begin{eqnarray}\label{detty2}
(L_1L_4-L_2L_3)\psi_\perp=0, 
\end{eqnarray}
\end{subequations}
where $(L_1L_4-L_2L_3)$ is the operational determinant of system (\ref{systemefinL}). By inspection, this determinant is non-zero, and thus a unique solution to system (\ref{systemefinL}) exists. We are free to solve Eq.~(\ref{detty1}) to obtain a solution for $\phi_\perp$ and then use either of Eqs.~(\ref{systemefinL1}) or (\ref{systemefinL2}) to solve for $\psi_\perp$, or alternatively, we can first solve Eq.~(\ref{detty2}) for $\psi_\perp$ and then use either of Eqs.~(\ref{systemefinL1}) or (\ref{systemefinL2}) to solve for $\phi_\perp$. For convenience, we choose the first option.  

After some straightforward algebra, it can be shown that Eq.~(\ref{detty1}) assumes the form
\begin{subequations}\label{phipde} 
\begin{equation}\label{phipde1}
\left[a_2\nabla_\perp^4+a_1\nabla_\perp^2+a_0\right]\phi_\perp=0,
\end{equation}
where
\begin{equation}\label{phipde2}
a_2=\mu(\lambda+2\mu),
\end{equation}
\begin{equation}\label{phipde3}
a_1=2\mu(\lambda+2\mu)\kappa-(\lambda+3\mu)\rho\tau,
\end{equation}
\begin{equation}\label{phipde4}
a_0=(\mu\kappa-\rho\tau)\left[(\lambda+2\mu)\kappa-\rho\tau\right].
\end{equation}
\end{subequations}
Since the operator in Eq.~(\ref{phipde1}) is a polynomial differential operator in $\nabla_\perp^2$ with constant coefficients, we may look for separable solutions (denoted here by $\varphi$) of the same fundamental form as those of the two-dimensional Helmholtz equation in polar coordinates ($\nabla_\perp^2\varphi=\Lambda\varphi$). This leads to the characteristic equation 
\begin{equation}\label{chareq}
a_2\Lambda^2+a_1\Lambda+a_0=0,
\end{equation}
whose roots, obtained from the quadratic formula
\begin{subequations}\label{rootsc} 
\begin{equation}\label{rootsc1}
\Lambda_{\pm}={-a_1\pm\sqrt{a_1^2-4a_2a_0}\over2a_2},
\end{equation}
are
\begin{equation}\label{rootsSC}
\Lambda_-=-\left(\kappa-{\rho\tau\over(\lambda+2\mu)}\right), \quad \Lambda_+=-\left(\kappa-{\rho\tau\over\mu}\right).
\end{equation}
\end{subequations}

\vspace*{0.25cm}

If $\varphi_-(r,\theta)$ and $\varphi_+(r,\theta)$ denote two independent separable solutions to Eq.~(\ref{phipde}) corresponding to the roots $\Lambda_-$ and $\Lambda_+$, respectively, then $\phi_\perp(r,\theta)=\varphi_-(r,\theta)+\varphi_+(r,\theta)$. 
We shall, for notational convenience, let $\phi^{(1)}_\perp(r,\theta)\equiv\varphi_-(r,\theta)$ and $\phi^{(2)}_\perp(r,\theta)\equiv\varphi_+(r,\theta)$. In this more intuitive notation, 
\begin{equation}\label{phiperpGEN}
\phi_\perp(r,\theta)=\sum_{s=1}^2\phi^{(s)}_\perp(r,\theta)=\sum_{s=1}^2\phi^{(s)}_r(r)\phi^{(s)}_\theta(\theta).
\end{equation}
The angular parts of $\varphi_-(r,\theta)$ and $\varphi_+(r,\theta)$ [i.e., $\phi^{(1)}_\theta(\theta)$ and $\phi^{(2)}_\theta(\theta)$ in Eq.~(\ref{phiperpGEN}), respectively] can be either periodic or non-periodic functions of $\theta$. In SC1 \cite{usB1}, we considered exclusively the class of separable solutions to system (\ref{systemefinL}) that possess $2\pi$-periodic angular parts. In the present paper, we shall consider the complementary class of separable solutions to system (\ref{systemefinL}) whose angular parts are either aperiodic or periodic but not $2\pi$-periodic. Such solutions are useful for solving a wide variety of dynamical problems involving open cylindrical shells and panels as well as open solid cylinders, where $2\pi$-periodicity in the angular coordinate is neither necessary nor (depending on the boundary conditions) appropriate \cite{panelEG1,panelEG2}. 

Utilizing the fundamental solutions to the 2D Helmholtz equation in polar coordinates (see Appendix \ref{twodimHelmPCs}) with the Helmholtz constant taking the values $\Lambda_1\equiv\Lambda_-$ and $\Lambda_2\equiv\Lambda_+$, we obtain the following particular solution to Eq.~(\ref{phipde}):   
\begin{subequations}\label{phiNP}
\begin{equation}
\phi_\perp(r,\theta)=\sum_{s=1}^2\phi^{(s)}_r(r)\phi^{(s)}_\theta(\theta),
\end{equation}
where
\begin{eqnarray}\label{phiangpartNP}
\phi^{(s)}_\theta(\theta)=\left \{ \begin{array}{lr}
             C_s\exp\left(-\sqrt{|\eta|}\theta\right)+D_s\exp\left(\sqrt{|\eta|}\theta\right) & \text{if}~\eta<0 \\
             C_s + D_s\theta & \text{if}~\eta=0 \\
             C_s\cos\left(\sqrt{\eta}\theta\right)+D_s\sin\left(\sqrt{\eta}\theta\right) & \text{if}~\eta>0
           \end{array} \right.,
\end{eqnarray}
\begin{eqnarray}\label{phiradsolNP}
\phi^{(s)}_r(r)=\left \{ \begin{array}{lr}
          A_sJ_{i\sqrt{|\eta|}}\left(\sqrt{|\Lambda_s|}r\right)+B_sY_{i\sqrt{|\eta|}}\left(\sqrt{|\Lambda_s|}r\right) & \text{if}~\Lambda_s<0~\text{and}~\eta<0 \\
            A_s\cos\left(\sqrt{|\eta|}\ln r\right)+B_s\sin\left(\sqrt{|\eta|}\ln r\right) & \text{if}~\Lambda_s=0~\text{and}~\eta<0 \\
            A_sI_{i\sqrt{|\eta|}}\left(\sqrt{\Lambda_s}r\right)+B_sK_{i\sqrt{|\eta|}}\left(\sqrt{\Lambda_s}r\right) & \text{if}~\Lambda_s>0~\text{and}~\eta<0 \\
           A_sJ_0\left(\sqrt{|\Lambda_s|}r\right)+B_sY_0\left(\sqrt{|\Lambda_s|}r\right) &  \text{if}~\Lambda_s<0~\text{and}~\eta=0 \\
           A_s+B_s\ln r & \text{if}~\Lambda_s=0~\text{and}~\eta=0 \\ 
            A_sI_0\left(\sqrt{\Lambda_s}r\right)+B_sK_0\left(\sqrt{\Lambda_s}r\right) & \text{if}~\Lambda_s>0~\text{and}~\eta=0 \\
          A_sJ_{\sqrt{\eta}}\left(\sqrt{|\Lambda_s|}r\right)+B_sY_{\sqrt{\eta}}\left(\sqrt{|\Lambda_s|}r\right) & \text{if}~\Lambda_s<0~\text{and}~\eta>0 \\ 
            A_sr^{\sqrt{\eta}}+B_sr^{-\sqrt{\eta}} & \text{if}~\Lambda_s=0~\text{and}~\eta>0 \\
            A_sI_{\sqrt{\eta}}\left(\sqrt{\Lambda_s}r\right)+B_sK_{\sqrt{\eta}}\left(\sqrt{\Lambda_s}r\right) & \text{if}~\Lambda_s>0~\text{and}~\eta>0 \\
           \end{array} \right.,
\end{eqnarray}  
\end{subequations}
and $\{A_1,A_2,B_1,B_2,C_1,C_2,D_1,D_2\}$ are arbitrary constants. Note that the above solution involves \emph{three} parameters: $\displaystyle \kappa\in\mathbb{R}\backslash\{0\}$ and $\displaystyle \tau\in\mathbb{R}\backslash\{0\}$, which occur in the (non-arbitrary) constants $\{\Lambda_1,\Lambda_2\}$, and $\eta\in\mathbb{R}$. The parameter $\eta$ in solution (\ref{phiNP}) is analogous to the constant $n\in\mathbb{N}$ that appears in the solution to Eq.~(\ref{phipde}) when the circumferential functions $\left\{\phi^{(s)}_\theta(\theta): s=1,2\right\}$ are $2\pi$-periodic \cite{usB1}. Indeed, it can be readily verified that solution (\ref{phiNP}) reduces to the analogous $2\pi$-periodic solution obtained in SC1 \cite{usB1} when $\eta\in\left\{n^2:n\in\mathbb{N}\right\}$ with the caveat that $D_s=0~(s=1,2)$ when $n=0$. Henceforth, we shall assume $\eta\in\mathbb{R}\backslash\left\{N^2:N\in\mathbb{Z}^+\right\}$.  

It is important to recognize that, when the parameter $\eta<0$, the radial functions $\left\{\phi^{(s)}_r(r): s=1,2\right\}$ are generally complex-valued functions. Real particular solutions can however be extracted in these cases. When $\Lambda_s>0$ and $\eta<0$, $\phi^{(s)}_r(r)$ is a solution to 
\begin{equation}\label{CS4NP}
r^2{\mathrm{d}^2\phi^{(s)}_r(r)\over\mathrm{d}r^2} + r{\mathrm{d}\phi^{(s)}_r(r)\over\mathrm{d}r}+\left(|\eta|-\Lambda_sr^2\right)\phi^{(s)}_r(r)=0,
\end{equation}
and based on the discussion in the first paragraph of Appendix \ref{APPBSSLIMG}, the \emph{real}-valued functions
\begin{equation}
\Bigg\{\text{Re}\left\{I_{i\sqrt{|\eta|}}\left(\sqrt{\Lambda_s}r\right)\right\},K_{i\sqrt{|\eta|}}\left(\sqrt{\Lambda_s}r\right)\Bigg\}
\end{equation}
constitute a linearly independent set of solutions to Eq.~(\ref{CS4NP}). When $\Lambda_s<0$ and $\eta<0$, $\phi^{(s)}_r(r)$ is a solution to 
\begin{equation}\label{CS2NP}
r^2{\mathrm{d}^2\phi^{(s)}_r(r)\over\mathrm{d}r^2} + r{\mathrm{d}\phi^{(s)}_r(r)\over\mathrm{d}r}+\left(|\Lambda_s|r^2+|\eta|\right)\phi^{(s)}_r(r)=0,
\end{equation}
and in this case (c.f., the discussion in the second paragraph of Appendix \ref{APPBSSLIMG}), the functions
\begin{subequations}
\begin{eqnarray}
\bar{J}_{\sqrt{|\eta|}}\left(\sqrt{|\Lambda_s|}r\right)&\equiv&\text{sech}\left({\pi\sqrt{|\eta|}\over2}\right)\text{Re}\left\{J_{i\sqrt{|\eta|}}\left(\sqrt{|\Lambda_s|}r\right)\right\}, \\
\overline{Y}_{\sqrt{|\eta|}}\left(\sqrt{|\Lambda_s|}r\right)&\equiv&\text{sech}\left({\pi\sqrt{|\eta|}\over2}\right)\text{Re}\left\{Y_{i\sqrt{|\eta|}}\left(\sqrt{|\Lambda_s|}r\right)\right\},
\end{eqnarray}
\end{subequations}
together constitute a \emph{real} and linearly independent set of solutions to Eq.~(\ref{CS2NP}). 

To obtain the solution for $\psi_\perp(r,\theta)$, it is simplest to re-arrange Eq.~(\ref{systemefinL1}) and (using the solution for $\phi_\perp(r,\theta)$) directly solve for $\psi_\perp(r,\theta)$: 
\begin{equation}\label{psi_sol_sp}
\psi_\perp(r,\theta)=-{\left[(\lambda+2\mu)\nabla_\perp^2+(\mu\kappa - \rho\tau)\right]\over(\lambda+\mu)\kappa}\phi_\perp(r,\theta).
\end{equation}
In SC1 \cite{usB1}, it is shown that, given the solution for $\phi_\perp(r,\theta)$ as expressed in Eq.~(\ref{phiperpGEN}), the solution for $\psi_\perp(r,\theta)$ can be expressed as: 
\begin{subequations}\label{solpsiPExp2}
\begin{equation}\label{solpsiPExp2A}
\psi_\perp(r,\theta)=\sum_{s=1}^2\gamma_s\phi^{(s)}_r(r)\phi^{(s)}_\theta(\theta),
\end{equation} 
where the constant $\gamma_s~(s=1,2)$ is given by 
\begin{eqnarray}\label{solpsiPP2}
\displaystyle
\gamma_s=\left \{ \begin{array}{lr}
             1 & \text{if}~s=1 \\
             {1\over\kappa}\left(\kappa - {\rho\tau\over\mu}\right) & \text{if}~s=2
           \end{array} \right.. 
\end{eqnarray}
\end{subequations}
The solution for $\psi_\perp(r,\theta)$ is thus given by Eq.~(\ref{solpsiPExp2}), where the functions $\left\{\phi^{(s)}_r(r): s=1,2\right\}$ and $\left\{\phi^{(s)}_\theta(\theta): s=1,2\right\}$ are now given by Eqs.~(\ref{phiradsolNP}) and (\ref{phiangpartNP}), respectively. 

The only remaining task in solving system (\ref{systemeF}) is to determine the axial and temporal parts of the full potentials $\Phi$ and $\Psi$ [see Eq.~(\ref{solform})]. The conditions listed in (\ref{assump}) provide the governing equations for determining these functions. Conditions (\ref{assump3}), (\ref{assump1}), and (\ref{assump2}) yield
\begin{eqnarray}\label{Zpart}
\phi_z(z)=\psi_z(z)=\left \{ \begin{array}{lr}
             E\cos\left(\sqrt{|\kappa|}z\right)+F\sin\left(\sqrt{|\kappa|}z\right) & \text{if}~\kappa<0 \\
             E\exp\left(-\sqrt{\kappa}z\right)+F\exp\left(\sqrt{\kappa}z\right) & \text{if}~\kappa>0
           \end{array} \right.,
\end{eqnarray}
and similarly, conditions (\ref{assump6}), (\ref{assump4}), and (\ref{assump5}) yield
\begin{eqnarray}\label{Tpart}
\phi_t(t)=\psi_t(t)=\left \{ \begin{array}{lr}
             G\cos\left(\sqrt{|\tau|}t\right)+H\sin\left(\sqrt{|\tau|}t\right) & \text{if}~\tau<0 \\
             G\exp\left(-\sqrt{\tau}t\right)+H\exp\left(\sqrt{\tau}t\right) & \text{if}~\tau>0
           \end{array} \right.,
\end{eqnarray}
where $\{E,F,G,H\}$ are arbitrary constants. 

Particular separable solutions to subsystem (\ref{systemeSFIN3}) can be obtained by assuming a product solution of the form
\begin{equation}\label{chiprodsol}
\chi(r,\theta,z,t)=\chi^{}_r(r)\chi^{}_\theta(\theta)\chi^{}_z(z)\chi^{}_t(t)
\end{equation}
and using separation of variables, as shown in Appendix \ref{AppSOV}. The separation constants $\{\upsilon_t,\upsilon_z,\upsilon_\theta,\upsilon_r\}$ defined in Appendix \ref{AppSOV} are a set of independent  parameters in their own right, limited only by relation (\ref{constsrel}). Thus, in theory, the full solution to system (\ref{systemeSFIN}) can involve \emph{six} independent parameters (e.g., $\{\kappa,\tau,\eta,\upsilon_t,\upsilon_z,\upsilon_\theta\}$). The number of parameters in the full solution can however be systematically reduced (yielding more compact particular solutions) by fixing one or more of the separation constants at prescribed value(s) that are linear combinations of the parameters $\{\kappa,\tau,\eta\}$. Although not necessary, it is convenient (and often useful) to eliminate the separation constants $\{\upsilon_t,\upsilon_z,\upsilon_\theta,\upsilon_r\}$ entirely by fixing their values according to the following prescription: 
\begin{equation}\label{sepconsts}
\upsilon_t={\rho\tau\over\mu}, \quad \quad \upsilon_z=\kappa, \quad \quad \upsilon_\theta=\eta, \quad \quad \upsilon_r=\upsilon_t-\upsilon_z=\left({\rho\tau\over\mu}-\kappa\right)\equiv\Lambda_2.
\end{equation} 
Upon doing so, one can immediately conclude (from inspection of the solutions given in Appendix \ref{AppSOV}) that, apart from arbitrary constants, $\chi^{}_z(z)$ and $\chi^{}_t(t)$ are identical to $\phi_z(z)$ and $\phi_t(t)$, respectively, and that $\chi^{}_\perp(r,\theta)\equiv\chi^{}_r(r)\chi^{}_\theta(\theta)$ is (apart from arbitrary constants) identical to the $s=2$ term of Eq.~(\ref{phiNP}). Explicitly: 
\begin{subequations}\label{solnforchiINPAPkapneqzero}
\begin{eqnarray}\label{chiradsolNPAPMN}
\chi^{}_r(r)=\left \{ \begin{array}{lr}
           A_3J_{i\sqrt{|\eta|}}\left(\sqrt{|\Lambda_2|}r\right)+B_3Y_{i\sqrt{|\eta|}}\left(\sqrt{|\Lambda_2|}r\right) & \text{if}~\Lambda_2<0~\text{and}~\eta<0 \\
            A_3\cos\left(\sqrt{|\eta|}\ln r\right)+B_3\sin\left(\sqrt{|\eta|}\ln r\right) & \text{if}~\Lambda_2=0~\text{and}~\eta<0 \\
           A_3I_{i\sqrt{|\eta|}}\left(\sqrt{\Lambda_2}r\right)+B_3K_{i\sqrt{|\eta|}}\left(\sqrt{\Lambda_2}r\right) & \text{if}~\Lambda_2>0~\text{and}~\eta<0 \\
           A_3J_0\left(\sqrt{|\Lambda_2|}r\right)+B_3Y_0\left(\sqrt{|\Lambda_2|}r\right) & \text{if}~\Lambda_2<0~\text{and}~\eta=0 \\
           A_3+B_3\ln r & \text{if}~\Lambda_2=0~\text{and}~\eta=0 \\ 
           A_3I_0\left(\sqrt{\Lambda_2}r\right)+B_3K_0\left(\sqrt{\Lambda_2}r\right) & \text{if}~\Lambda_2>0~\text{and}~\eta=0 \\
           A_3J_{\sqrt{\eta}}\left(\sqrt{|\Lambda_2|}r\right)+B_3Y_{\sqrt{\eta}}\left(\sqrt{|\Lambda_2|}r\right) & \text{if}~\Lambda_2<0~\text{and}~\eta>0 \\ 
            A_3r^{\sqrt{\eta}}+B_3r^{-\sqrt{\eta}} & \text{if}~\Lambda_2=0~\text{and}~\eta>0 \\
            A_3I_{\sqrt{\eta}}\left(\sqrt{\Lambda_2}r\right)+B_3K_{\sqrt{\eta}}\left(\sqrt{\Lambda_2}r\right) & \text{if}~\Lambda_2>0~\text{and}~\eta>0 \\
           \end{array} \right., 
\end{eqnarray} 
\begin{eqnarray}\label{chiangpartMN}
\chi^{}_\theta(\theta)=\left \{ \begin{array}{lr}
             C_3\exp\left(-\sqrt{|\eta|}\theta\right)+D_3\exp\left(\sqrt{|\eta|}\theta\right) & \text{if}~\eta<0 \\
             C_3 + D_3\theta & \text{if}~\eta=0 \\
             C_3\cos\left(\sqrt{\eta}\theta\right)+D_3\sin\left(\sqrt{\eta}\theta\right) & \text{if}~\eta>0
           \end{array} \right.,
\end{eqnarray} 
\begin{eqnarray}\label{chiZsolMN}
\chi^{}_z(z)=\left \{ \begin{array}{lr}
             \widetilde{E}\cos\left(\sqrt{|\kappa|}z\right)+\widetilde{F}\sin\left(\sqrt{|\kappa|}z\right) & \text{if}~\kappa<0 \\ 
             \widetilde{E}\exp\left(-\sqrt{\kappa}z\right)+\widetilde{F}\exp\left(\sqrt{\kappa}z\right) & \text{if}~\kappa>0
           \end{array} \right.,
\end{eqnarray}  
and
\begin{eqnarray}\label{chiTsolMN}
\chi^{}_t(t)=\left \{ \begin{array}{lr}
             \widetilde{G}\cos\left(\sqrt{|\tau|}t\right)+\widetilde{H}\sin\left(\sqrt{|\tau|}t\right) & \text{if}~\tau<0 \\ 
             \widetilde{G}\exp\left(-\sqrt{\tau}t\right)+\widetilde{H}\exp\left(\sqrt{\tau}t\right) & \text{if}~\tau>0
           \end{array} \right.,
\end{eqnarray} 
\end{subequations} 
where $\displaystyle\left\{A_3,B_3,C_3,D_3,\widetilde{E},\widetilde{F},\widetilde{G},\widetilde{H}\right\}$ are arbitrary constants. The function $\chi^{}_\theta(\theta)$ given by Eq.~(\ref{chiangpartMN}) is $2\pi$-periodic when $\eta\in\left\{N^2:N\in\mathbb{Z}^+\right\}$ and in the special case $\eta=D_3=0$. 

Since all three Buchwald potentials have now been completely determined (under conditions (\ref{assump}) and (\ref{sepconsts})), we can write down the corresponding particular solutions to Eq.~(\ref{NLE}) using the component form of Eq.~(\ref{Buchy}):
\begin{equation}\label{Buchy22}
\mathbf{u}=u_r\mathbf{\hat{r}}+u_\theta{\boldsymbol{\hat{\theta}}}+u_z\mathbf{\hat{z}}=\left({\partial\Phi\over\partial r}+{1\over r}{\partial\chi\over\partial\theta}\right)\mathbf{\hat{r}} + \left({1\over r}{\partial\Phi\over\partial\theta}-{\partial\chi\over\partial r}\right){\boldsymbol{\hat{\theta}}}+{\partial\Psi\over\partial z}\mathbf{\hat{z}}. 
\end{equation}     
Using Eq.~(\ref{Buchy22}), the cylindrical components of the displacement field are thus given by:
\begin{subequations}\label{DISCOMPN2PPkapdiffzero}
\begin{equation}\label{URNP}
u_r=\left(\sum_{s=1}^2{\text{d}\phi^{(s)}_r(r)\over\text{d}r}\phi^{(s)}_\theta(\theta)\right)\phi_z(z)\phi_t(t)+{1\over r}\chi^{}_r(r){\text{d}\chi^{}_\theta(\theta)\over\text{d}\theta}\chi^{}_z(z)\chi^{}_t(t),
\end{equation}
\begin{equation}\label{UTHNP}
u_\theta={1\over r}\left(\sum_{s=1}^2\phi^{(s)}_r(r){\text{d}\phi^{(s)}_\theta(\theta)\over\text{d}\theta}\right)\phi_z(z)\phi_t(t)-{\text{d}\chi^{}_r(r)\over\text{d}r}\chi^{}_\theta(\theta)\chi^{}_z(z)\chi^{}_t(t),
\end{equation}
\begin{equation}\label{UZNP}
u_z=\left(\sum_{s=1}^2\gamma_s\phi^{(s)}_r(r)\phi^{(s)}_\theta(\theta)\right){\text{d}\psi_z(z)\over\text{d}z}\psi_t(t),
\end{equation}
\end{subequations}
where:

\noindent \textbf{(i)} $\phi^{(s)}_r(r)$ and $\phi^{(s)}_\theta(\theta)$ in Eqs.~(\ref{URNP})-(\ref{UZNP}) are given by Eqs.~(\ref{phiradsolNP}) and (\ref{phiangpartNP}), respectively;

\noindent \textbf{(ii)} $\phi_z(z)$ and $\phi_t(t)$ in Eqs.~(\ref{URNP})-(\ref{UTHNP}) are given by Eqs.~(\ref{Zpart}) and (\ref{Tpart}), respectively;

\noindent \textbf{(iii)} $\chi^{}_r(r)$ and $\chi^{}_\theta(\theta)$ in Eqs.~(\ref{URNP})-(\ref{UTHNP}) are given by Eqs.~(\ref{chiradsolNPAPMN}) and (\ref{chiangpartMN}), respectively;

\noindent \textbf{(iv)} $\chi^{}_z(z)$ and $\chi^{}_t(t)$ in Eqs.~(\ref{URNP})-(\ref{UTHNP}) are given by Eqs.~(\ref{chiZsolMN}) and (\ref{chiTsolMN}), respectively;

\noindent \textbf{(v)} $\psi_z(z)$ and $\psi_t(t)$ in Eq.~(\ref{UZNP}) are given by Eqs.~(\ref{Zpart}) and (\ref{Tpart}), respectively;

\noindent \textbf{(vi)} the constant $\gamma_s$ in Eq.~(\ref{UZNP}) is given by Eq.~(\ref{solpsiPP2}).  

\noindent \textbf{Note:} Items \textbf{(iii)} and \textbf{(iv)} above are valid only under prescription (\ref{sepconsts}). If and when prescription (\ref{sepconsts}) is inapplicable (as in the example problem considered in Sec.~\ref{EGFULL}), then the functions $\left\{\chi^{}_r(r),\chi^{}_\theta(\theta),\chi^{}_z(z),\chi^{}_t(t)\right\}$ in Eqs.~(\ref{URNP})-(\ref{UTHNP}) are given by Eqs.~(\ref{chiangpart})-(\ref{chiradsolNPAP}) of Appendix \ref{AppSOV}.  

\subsubsection{Special $2\pi$-Periodic Subsolution: No Dependence on the Circumferential Coordinate $\displaystyle\theta$} 

When the circumferential functions $\left\{\phi^{(s)}_\theta(\theta): s=1,2\right\}$ and $\chi^{}_\theta(\theta)$ are constant functions, the Buchwald potentials (and the resulting displacement field) are independent of the $\theta$ coordinate, and Eqs.~(\ref{URNP})-(\ref{UZNP}) reduce to the following general form: 
\begin{subequations}\label{DISCOMPeg}
\begin{eqnarray}
u_r&=&\left(\sum_{s=1}^2C_s{\text{d}\phi^{(s)}_r(r)\over\text{d}r}\right)\phi_z(z)\phi_t(t), \label{URNPeg} \\
u_\theta&=&-C_3{\text{d}\chi^{}_r(r)\over\text{d}r}\chi^{}_z(z)\chi^{}_t(t), \label{UTHNPeg} \\
u_z&=&\left(\sum_{s=1}^2\gamma_sC_s\phi^{(s)}_r(r)\right){\text{d}\psi_z(z)\over\text{d}z}\psi_t(t). \label{UZNPeg}
\end{eqnarray}
\end{subequations}
In this case, the Buchwald potentials (and the resulting displacement field) are trivially $2\pi$-periodic. It should however be emphasized that the absence of the $\theta$ coordinate in the solution does not mean that the displacement field is necessarily axisymmetric. Axisymmetric displacement fields must, in addition, have no circumferential component (i.e., $u_\theta=0$). An example of an \emph{asymmetric} problem having no explicit dependence on the $\theta$ coordinate is given in Sec.~\ref{EGFULL}.         

\subsection{Special Case: $\displaystyle \kappa=0$}

When the parameter $\kappa=0$, system (\ref{systemefin}) decouples and reduces to: 
\begin{subequations}\label{systemefinkapeq0} 
\begin{equation}\label{systemefin1kapeq0}
\left[(\lambda+2\mu)\nabla_\perp^2 - \rho\tau\right]\phi_\perp=0,
\end{equation}
\begin{equation}\label{systemefin2kapeq0}
\left[(\lambda+\mu)\nabla_\perp^2\right]\phi_\perp+\left[\mu\nabla_\perp^2-\rho\tau\right]\psi_\perp=0. 
\end{equation}
\end{subequations}
Equation (\ref{systemefin1kapeq0}) can be cast in the form of a 2D Helmholtz equation (in polar coordinates) and immediately solved using separation of variables. Letting $\Lambda^{(0)}_1\equiv{\rho\tau\over(\lambda+2\mu)}$, the following separable solution to Eq.~(\ref{systemefin1kapeq0}) is then obtained using the results of Appendix \ref{twodimHelmPCs}: 
\begin{subequations}\label{solphiPkapeq0}
\begin{equation}\label{solphiPkapeq0A}
\phi_\perp(r,\theta)=R_1(r)\Theta_1(\theta),
\end{equation}
where
\begin{eqnarray}\label{solphiPkapeq0B}
\Theta_1(\theta)=\left \{ \begin{array}{lr}
             C_1\exp\left(-\sqrt{|\eta|}\theta\right)+D_1\exp\left(\sqrt{|\eta|}\theta\right) & \text{if}~\eta<0 \\
             C_1 + D_1\theta & \text{if}~\eta=0 \\
             C_1\cos\left(\sqrt{\eta}\theta\right)+D_1\sin\left(\sqrt{\eta}\theta\right) & \text{if}~\eta>0
           \end{array} \right.,
\end{eqnarray}
\begin{eqnarray}\label{solphiPkapeq0C}
R_1(r)=\left \{ \begin{array}{lr}
          A_1J_{i\sqrt{|\eta|}}\left(\sqrt{\left|\Lambda^{(0)}_1\right|}r\right)+B_1Y_{i\sqrt{|\eta|}}\left(\sqrt{\left|\Lambda^{(0)}_1\right|}r\right) & \text{if}~\tau<0~\text{and}~\eta<0 \\
          A_1I_{i\sqrt{|\eta|}}\left(\sqrt{\Lambda^{(0)}_1}r\right)+B_1K_{i\sqrt{|\eta|}}\left(\sqrt{\Lambda^{(0)}_1}r\right) & \text{if}~\tau>0~\text{and}~\eta<0 \\
          A_1J_0\left(\sqrt{\left|\Lambda^{(0)}_1\right|}r\right)+B_1Y_0\left(\sqrt{\left|\Lambda^{(0)}_1\right|}r\right) &  \text{if}~\tau<0~\text{and}~\eta=0 \\
          A_1I_0\left(\sqrt{\Lambda^{(0)}_1}r\right)+B_1K_0\left(\sqrt{\Lambda^{(0)}_1}r\right) & \text{if}~\tau>0~\text{and}~\eta=0 \\
          A_1J_{\sqrt{\eta}}\left(\sqrt{\left|\Lambda^{(0)}_1\right|}r\right)+B_1Y_{\sqrt{\eta}}\left(\sqrt{\left|\Lambda^{(0)}_1\right|}r\right) & \text{if}~\tau<0~\text{and}~\eta>0 \\ 
          A_1I_{\sqrt{\eta}}\left(\sqrt{\Lambda^{(0)}_1}r\right)+B_1K_{\sqrt{\eta}}\left(\sqrt{\Lambda^{(0)}_1}r\right) & \text{if}~\tau>0~\text{and}~\eta>0 \\
           \end{array} \right., 
\end{eqnarray}  
\end{subequations}
where $\{A_1,B_1,C_1,D_1\}$ are arbitrary constants. 

To solve for $\psi_\perp(r,\theta)$, we substitute $(\lambda+\mu)\nabla_\perp^2\phi_\perp=-\left[\mu\nabla_\perp^2 - \rho\tau\right]\phi_\perp$ (which is obtained from rearranging Eq.~(\ref{systemefin1kapeq0})) into Eq.~(\ref{systemefin2kapeq0}). This yields the PDE: 
\begin{equation}\label{FperpeqA}
\left[\mu\nabla_\perp^2 - \rho\tau\right]\mathsf{W}_\perp=0,~~\text{where}~~\mathsf{W}_\perp\equiv\left(\psi_\perp-\phi_\perp\right).
\end{equation}
As before, Eq.~(\ref{FperpeqA}) can be cast in the form of a 2D Helmholtz equation in polar coordinates. 
Letting $\Lambda^{(0)}_2\equiv{\rho\tau\over\mu}$, the following separable solution to Eq.~(\ref{FperpeqA}) is then obtained using the results of Appendix \ref{twodimHelmPCs}:
\begin{subequations}\label{solFperpeq}
\begin{equation}\label{solFperpeqA}
\mathsf{W}_\perp(r,\theta)=R_2(r)\Theta_2(\theta),  
\end{equation}
where
\begin{eqnarray}\label{solFperpeqB}
\Theta_2(\theta)=\left \{ \begin{array}{lr}
             C_2\exp\left(-\sqrt{|\eta|}\theta\right)+D_2\exp\left(\sqrt{|\eta|}\theta\right) & \text{if}~\eta<0 \\
             C_2 + D_2\theta & \text{if}~\eta=0 \\
             C_2\cos\left(\sqrt{\eta}\theta\right)+D_2\sin\left(\sqrt{\eta}\theta\right) & \text{if}~\eta>0
           \end{array} \right.,
\end{eqnarray}
\begin{eqnarray}\label{solFperpeqC}
R_2(r)=\left \{ \begin{array}{lr}
          A_2J_{i\sqrt{|\eta|}}\left(\sqrt{\left|\Lambda^{(0)}_2\right|}r\right)+B_2Y_{i\sqrt{|\eta|}}\left(\sqrt{\left|\Lambda^{(0)}_2\right|}r\right) & \text{if}~\tau<0~\text{and}~\eta<0 \\
          A_2I_{i\sqrt{|\eta|}}\left(\sqrt{\Lambda^{(0)}_2}r\right)+B_2K_{i\sqrt{|\eta|}}\left(\sqrt{\Lambda^{(0)}_2}r\right) & \text{if}~\tau>0~\text{and}~\eta<0 \\
          A_2J_0\left(\sqrt{\left|\Lambda^{(0)}_2\right|}r\right)+B_2Y_0\left(\sqrt{\left|\Lambda^{(0)}_2\right|}r\right) &  \text{if}~\tau<0~\text{and}~\eta=0 \\
          A_2I_0\left(\sqrt{\Lambda^{(0)}_2}r\right)+B_2K_0\left(\sqrt{\Lambda^{(0)}_2}r\right) & \text{if}~\tau>0~\text{and}~\eta=0 \\
          A_2J_{\sqrt{\eta}}\left(\sqrt{\left|\Lambda^{(0)}_2\right|}r\right)+B_2Y_{\sqrt{\eta}}\left(\sqrt{\left|\Lambda^{(0)}_2\right|}r\right) & \text{if}~\tau<0~\text{and}~\eta>0 \\ 
          A_2I_{\sqrt{\eta}}\left(\sqrt{\Lambda^{(0)}_2}r\right)+B_2K_{\sqrt{\eta}}\left(\sqrt{\Lambda^{(0)}_2}r\right) & \text{if}~\tau>0~\text{and}~\eta>0 \\
           \end{array} \right.,
\end{eqnarray}
\end{subequations}
where $\{A_2,B_2,C_2,D_2\}$ are arbitrary constants. 
Since $\mathsf{W}_\perp\equiv\left(\psi_\perp-\phi_\perp\right)$, the solution for $\psi_\perp(r,\theta)$ is then the sum of the two solutions given by Eqs.~(\ref{solphiPkapeq0}) and (\ref{solFperpeq}):
\begin{equation}\label{solpsiperpkapeq0}
\psi_\perp(r,\theta)=\sum_{s=1}^2R_s(r)\Theta_s(\theta).   
\end{equation}
The axial parts of the full potentials $\Phi$ and $\Psi$ [c.f., Eqs.~(\ref{solform})], as determined from conditions (\ref{assump3}), (\ref{assump1}), and (\ref{assump2}) are: 
\begin{equation}\label{Zpartkapeq0}
\phi_z(z)=\psi_z(z)=E+Fz, 
\end{equation}
where $\{E,F\}$ are arbitrary constants. The temporal parts of potentials $\Phi$ and $\Psi$ are unchanged [c.f., Eq.~(\ref{Tpart})]. 

All that remains is to determine the Buchwald potential $\chi$, which is governed by Eq.~(\ref{systemeSFIN3}), and by assumption, has the separable form (\ref{chiprodsol}). Under conditions (\ref{sepconsts}), one can immediately conclude from inspection of the solutions given in Appendix \ref{AppSOV} that, apart from arbitrary constants, the components $\chi^{}_z(z)$ and $\chi^{}_t(t)$ are identical to $\phi_z(z)$ and $\phi_t(t)$ as given by Eqs.~(\ref{Zpartkapeq0}) and (\ref{Tpart}), respectively, and that $\chi^{}_\perp(r,\theta)\equiv\chi^{}_r(r)\chi^{}_\theta(\theta)$ is (apart from arbitrary constants) identical to the function $\mathsf{W}_\perp(r,\theta)$ given by Eq.~(\ref{solFperpeq}). Thus, analogous to Eqs.~(\ref{chiradsolNPAPMN})-(\ref{chiZsolMN}), the spatial components of $\chi$ are as follows:  
\begin{subequations}\label{solchiperpkapeq0}
\begin{eqnarray}\label{solchiperpkapeq0A}
\chi^{}_\theta(\theta)=\left \{ \begin{array}{lr}
             C_3\exp\left(-\sqrt{|\eta|}\theta\right)+D_3\exp\left(\sqrt{|\eta|}\theta\right) & \text{if}~\eta<0 \\
             C_3 + D_3\theta & \text{if}~\eta=0 \\
             C_3\cos\left(\sqrt{\eta}\theta\right)+D_3\sin\left(\sqrt{\eta}\theta\right) & \text{if}~\eta>0
           \end{array} \right.,
\end{eqnarray}
\begin{eqnarray}\label{solchiperpkapeq0B}
\chi^{}_r(r)=\left \{ \begin{array}{lr}
          A_3J_{i\sqrt{|\eta|}}\left(\sqrt{\left|\Lambda^{(0)}_2\right|}r\right)+B_3Y_{i\sqrt{|\eta|}}\left(\sqrt{\left|\Lambda^{(0)}_2\right|}r\right) & \text{if}~\tau<0~\text{and}~\eta<0 \\
          A_3I_{i\sqrt{|\eta|}}\left(\sqrt{\Lambda^{(0)}_2}r\right)+B_3K_{i\sqrt{|\eta|}}\left(\sqrt{\Lambda^{(0)}_2}r\right) & \text{if}~\tau>0~\text{and}~\eta<0 \\
          A_3J_0\left(\sqrt{\left|\Lambda^{(0)}_2\right|}r\right)+B_3Y_0\left(\sqrt{\left|\Lambda^{(0)}_2\right|}r\right) &  \text{if}~\tau<0~\text{and}~\eta=0 \\
          A_3I_0\left(\sqrt{\Lambda^{(0)}_2}r\right)+B_3K_0\left(\sqrt{\Lambda^{(0)}_2}r\right) & \text{if}~\tau>0~\text{and}~\eta=0 \\
          A_3J_{\sqrt{\eta}}\left(\sqrt{\left|\Lambda^{(0)}_2\right|}r\right)+B_3Y_{\sqrt{\eta}}\left(\sqrt{\left|\Lambda^{(0)}_2\right|}r\right) & \text{if}~\tau<0~\text{and}~\eta>0 \\ 
          A_3I_{\sqrt{\eta}}\left(\sqrt{\Lambda^{(0)}_2}r\right)+B_3K_{\sqrt{\eta}}\left(\sqrt{\Lambda^{(0)}_2}r\right) & \text{if}~\tau>0~\text{and}~\eta>0 \\
           \end{array} \right.,
\end{eqnarray}
\begin{equation}\label{ZpartCkapeq0}
\chi^{}_z(z)=\widetilde{E}+\widetilde{F}z, 
\end{equation}
\end{subequations}
where $\left\{A_3,B_3,C_3,D_3,\widetilde{E},\widetilde{F}\right\}$ are arbitrary constants. The temporal part of $\chi$ (i.e., $\chi^{}_t(t)$) is again given by Eq.~(\ref{chiTsolMN}).  

With all three Buchwald potentials now completely determined (under the stipulated conditions), we can write down the corresponding particular solutions to Eq.~(\ref{NLE}) using (\ref{Buchy22}):
\begin{subequations}\label{finsolnkapeq0}
\begin{equation}\label{solnradcompkapeq0}
u_r={\text{d}R_1(r)\over\text{d}r}\Theta_1(\theta)\phi_z(z)\phi_t(t)+{1\over r}\chi^{}_r(r){\text{d}\chi^{}_\theta(\theta)\over\text{d}\theta}\chi^{}_z(z)\chi^{}_t(t),
\end{equation}
\begin{equation}\label{solnangcompkapeq0}
u_\theta={1\over r}R_1(r){\text{d}\Theta_1(\theta)\over\text{d}\theta}\phi_z(z)\phi_t(t)-{\text{d}\chi^{}_r(r)\over\text{d}r}\chi^{}_\theta(\theta)\chi^{}_z(z)\chi^{}_t(t),
\end{equation}
\begin{equation}\label{solnaxcompkapeq0}
u_z=\left(\sum_{s=1}^2R_s(r)\Theta_s(\theta)\right){\text{d}\psi_z(z)\over\text{d}z}\psi_t(t),
\end{equation}
\end{subequations}
where: 

\noindent \textbf{(i)} $R_1(r)$ and $R_2(r)$ in Eqs.~(\ref{solnradcompkapeq0})-(\ref{solnaxcompkapeq0}) are given by Eqs.~(\ref{solphiPkapeq0C}) and (\ref{solFperpeqC}), respectively;

\noindent \textbf{(ii)} $\Theta_1(\theta)$ and $\Theta_2(\theta)$ in Eqs.~(\ref{solnradcompkapeq0})-(\ref{solnaxcompkapeq0}) are given by Eqs.~(\ref{solphiPkapeq0B}) and (\ref{solFperpeqB}), respectively; 

\noindent \textbf{(iii)} $\phi_z(z)$ and $\phi_t(t)$ in Eqs.~(\ref{solnradcompkapeq0})-(\ref{solnangcompkapeq0}) are given by Eqs.~(\ref{Zpartkapeq0}) and (\ref{Tpart}), respectively;

\noindent \textbf{(iv)} $\psi_z(z)$ and $\psi_t(t)$ in Eq.~(\ref{solnaxcompkapeq0}) are given by Eqs.~(\ref{Zpartkapeq0}) and (\ref{Tpart}), respectively; 

\noindent \textbf{(v)} $\chi^{}_r(r)$, $\chi^{}_\theta(\theta)$, $\chi^{}_z(z)$, and $\chi^{}_t(t)$ in Eqs.~(\ref{solnradcompkapeq0})-(\ref{solnangcompkapeq0}) are given by (\ref{solchiperpkapeq0B}), (\ref{solchiperpkapeq0A}), (\ref{ZpartCkapeq0}), and (\ref{chiTsolMN}), respectively. 

\noindent \textbf{Note 1:} Solution (\ref{finsolnkapeq0}) is valid only when the parameter $\kappa$ defined in Eqs.~(\ref{assump1})-(\ref{assump2}) is zero.  

\noindent \textbf{Note 2:} Item \textbf{(v)} above is valid only under prescription (\ref{sepconsts}). If and when prescription (\ref{sepconsts}) is inapplicable, then the functions $\left\{\chi^{}_r(r),\chi^{}_\theta(\theta),\chi^{}_z(z),\chi^{}_t(t)\right\}$ in Eqs.~(\ref{solnradcompkapeq0})-(\ref{solnaxcompkapeq0}) are given by Eqs.~(\ref{chiangpart})-(\ref{chiradsolNPAP}) of Appendix \ref{AppSOV}. 

\section{Applications}\label{AppPrologue}

Our focus thus far has been on solving the PDE system (\ref{systemeSFIN}) under assumptions (\ref{solform}) and (\ref{chiprodsol}) and conditions (\ref{assump}) and (\ref{sepconsts}) and then using (\ref{Buchy22}) to arrive at solutions (\ref{DISCOMPN2PPkapdiffzero}) and (\ref{finsolnkapeq0}). Independent of how these solutions were actually derived, we have at our disposal a set of particular solutions to the NL equation that can be directly applied to a fundamental set of linear-elastic boundary-value problems (BVPs) in cylindrical coordinates, namely, problems wherein the displacement, stress, and strain fields are: constant, linear, sinusoidal, exponential, or hyperbolic in the circumferential and longitudinal coordinates\footnote{Note that sinusoidal circumferential variations need not be $2\pi$-periodic.}, and sinusoidal or exponential in the time coordinate. Such solutions are useful for solving dynamical problems involving cylindrical geometries where (rational) $2\pi$-periodicity in the angular coordinate is incompatible with the given boundary conditions. In the remaining sections of the paper, we shall illustrate using simple (but mathematically non-trivial) examples how solutions (\ref{DISCOMPN2PPkapdiffzero}) and (\ref{finsolnkapeq0}) can be used to solve such BVPs. As an addendum to SC1, we first however present an example that illustrates the use of the special $2\pi$-periodic solution (\ref{DISCOMPeg}). This example is conceptually useful since it also furnishes an explicit example of a situation where prescription (\ref{sepconsts}) is inapplicable and one must use the more general solution for the Buchwald potential $\chi$ given in Appendix \ref{AppSOV}.  

\section{Application 1: Forced Vibration of a Closed Solid Cylinder}\label{EGFULL}

Consider a closed solid elastic circular cylinder of finite length $L$ and radius $R$ subjected to the following harmonic stresses on its curved surface:  
\begin{subequations}\label{BCsCRVD1}
\begin{equation}\label{strssrr1}
\sigma_{rr}(R,\theta,z,t)=\mathcal{A}\sin\left({k\pi\over L}z\right)\sin(\omega t),
\end{equation}
\begin{equation}\label{strssrt1}
\sigma_{r\theta}(R,\theta,z,t)=\mathcal{B}\sin\left({m\pi\over L}z\right),~~~~~~~~~
\end{equation}
\begin{equation}\label{strssrz1}
\sigma_{rz}(R,\theta,z,t)=\mathcal{C}\cos\left({k\pi\over L}z\right)\sin(\omega t),
\end{equation}
\end{subequations}
where $\sigma_{rr}(r,\theta,z,t)$ is a normal component of stress, $\sigma_{r\theta}(r,\theta,z,t)$ and $\sigma_{rz}(r,\theta,z,t)$ are shear components of stress, $\{\mathcal{A},\mathcal{B},\mathcal{C}\}$ are prescribed constant stresses, $\{k,m\}\in\mathbb{Z}^+$ are prescribed dimensionless constants, and $\omega$ is the prescribed angular frequency of excitation. Note that the circumferential stress given by Eq.~(\ref{strssrt1}) is time independent. The cylinder is simply-supported at its flat ends, which are situated at $z=0$ and $z=L$; the corresponding boundary conditions at the flat ends of the cylinder are thus:
\begin{subequations}\label{SSBCs}
\begin{equation}\label{SSBCs1}
u_r(r,\theta,0,t)=u_r(r,\theta,L,t)=0,
\end{equation}
\begin{equation}\label{SSBCs2}
u_\theta(r,\theta,0,t)=u_\theta(r,\theta,L,t)=0,
\end{equation}
\begin{equation}\label{SSBCs3}
\sigma_{zz}(r,\theta,0,t)=\sigma_{zz}(r,\theta,L,t)=0,
\end{equation}
\end{subequations}
where $\sigma_{zz}(r,\theta,z,t)$ is the normal component of stress along the axis of the cylinder. Conditions (\ref{BCsCRVD1}) are pure stress boundary conditions on the curved surface of the cylinder and must be satisfied for all $\theta\in[0,2\pi]$, $z\in(0,L)$, and arbitrary $t$.  Conditions (\ref{SSBCs}) are admissible mixed boundary conditions specifying orthogonal components of the stress and displacement at the flat ends of the cylinder; they must be satisfied for all $r\in[0,R]$, $\theta\in[0,2\pi]$, and arbitrary $t$. The condition that the displacement field is finite everywhere in the cylinder is implicit. 

For forced motion, \emph{at least} one of $\{\mathcal{A},\mathcal{C}\}$ must be non-zero. (When $\mathcal{A}=\mathcal{C}=0$, the cylinder does not vibrate.) Since the stresses are independent of the circumferential coordinate $\theta$ in this problem, the displacement field of the cylinder must be so as well. Note however that when \emph{all} of the stress constants $\{\mathcal{A},\mathcal{B},\mathcal{C}\}$ are non-zero, the motion is neither torsional (since both $\sigma_{rr}$ and $\sigma_{rz}$ are non-zero) nor axisymmetric (since $\sigma_{r\theta}$ is non-zero). 

\begin{framed}
\noindent {\textbf{Problem S}: In the special case where the angular excitation frequency $\omega=\sqrt{\lambda+2\mu\over\rho}\left({k\pi\over L}\right)$, determine the displacement field at all points of the simply-supported cylinder subject to the non-uniform conditions (\ref{BCsCRVD1}) on its curved surface.} 
\end{framed}

The requirement of $2\pi$-periodicity (since the cylinder is closed) and the fact that the displacement field is independent of $\theta$ (as deduced above) call upon the use of the special solution (\ref{DISCOMPeg}). 
Anticipating that the displacement component $u_\theta$ will be time independent (due to the time independence of the stress component $\sigma_{r\theta}$) and that the displacement components $u_r$ and $u_z$ must be time dependent implies that $\chi^{}_t(t)$ in Eq.~(\ref{UTHNPeg}) will not be of the same functional form as $\phi_t(t)$ and $\psi_t(t)$ in Eqs.~(\ref{URNPeg}) and (\ref{UZNPeg}), respectively. Furthermore, unless the constants $k$ and $m$ in (\ref{BCsCRVD1}) are equal (which they need not be), $\chi^{}_z(z)$ must be of a different functional form than $\phi_z(z)$ and $\psi_z(z)$. In such a situation, prescription (\ref{sepconsts}) and the solution for the Buchwald potential $\chi$ that results from it [i.e., Eqs.~(\ref{chiradsolNPAPMN})-(\ref{chiTsolMN})] are not valid and one must use the more general solution for $\chi$ given in Appendix \ref{AppSOV}. The parameters associated with the solution for $\chi$ (i.e., the separation constants $\{\upsilon_t,\upsilon_z,\upsilon_\theta,\upsilon_r\}$) must therefore be prescribed independently of the parameters associated with the solutions for $\Phi$ and $\Psi$ (i.e., $\{\kappa,\tau,\eta\}$). 

To obtain the general solutions for the $u_r$ and $u_z$ displacement components, fix the associated parameters $\kappa$, $\tau$, and $\eta$ as follows: 
\begin{equation}\label{kapandtau}
\kappa=-\left({k\pi\over L}\right)^2, \quad \tau=-\omega^2, \quad \eta=0. 
\end{equation}
Then, choosing $E=0$ and $F=1$ in Eq.~(\ref{Zpart}) and $G=0$ and $H=1$ in Eq.~(\ref{Tpart}) yields:
\begin{equation}\label{phipsiZEG}
\phi_z(z)=\psi_z(z)=\sin\left({k\pi\over L}z\right), 
\end{equation}
\begin{equation}\label{phipsiTEG}
\phi_t(t)=\psi_t(t)=\sin(\omega t).
\end{equation}
Choosing $C_1=C_2=1$ in Eq.~(\ref{phiangpartNP}) yields (noting that $D_1=D_2=\eta=0$ is required for $2\pi$-periodicity):
\begin{equation}\label{phithetassolcyl}
\phi_\theta^{(1)}(\theta)=\phi_\theta^{(2)}(\theta)=1. 
\end{equation}
The constants $\Lambda_1$ and $\Lambda_2$, as determined from Eq.~(\ref{rootsSC}), are:
\begin{equation}\label{LambdasEG}
\Lambda_1=\left({k\pi\over L}\right)^2-{\rho\omega^2\over(\lambda+2\mu)}=0, \quad \Lambda_2=\left({k\pi\over L}\right)^2-{\rho\omega^2\over\mu}<0.
\end{equation}
Since $\eta=0$, $\Lambda_1=0$, and $\Lambda_2<0$, it then follows from Eq.~(\ref{phiradsolNP}) that
\begin{equation}\label{phirssolcyl}
\phi_r^{(1)}(r)=A_1, \quad \phi_r^{(2)}(r)=A_2J_0\left(\sqrt{|\Lambda_2|}r\right), 
\end{equation}
where the $\ln r$ term generally present in the function $\phi_r^{(1)}(r)$ and the $Y_0(\sqrt{|\Lambda_2|}r)$ term generally present in the function $\phi_r^{(2)}(r)$ have been discarded (by prescribing $B_1=B_2=0$ in Eq.~(\ref{phiradsolNP})) in order to satisfy the finiteness condition at $r=0$. 

To obtain the general solution for $u_\theta$, fix the parameters $\upsilon_z$, $\upsilon_t$, and $\upsilon_\theta$ associated with the solution for $\chi$ as follows: 
\begin{equation}\label{upZTTH}
\upsilon_z=-\left({m\pi\over L}\right)^2, \quad \upsilon_t=0, \quad \upsilon_\theta=0. 
\end{equation}
The remaining parameter $\upsilon_r$ is then fixed by constraint (\ref{constsrel}): 
\begin{equation}\label{upR}
\upsilon_r=\upsilon_t-\upsilon_z=\left({m\pi\over L}\right)^2>0.
\end{equation}
Then, choosing $\widetilde{E}=0$ and $\widetilde{F}=1$ in Eq.~(\ref{chiZsol}) and $\widetilde{G}=1$ and $\widetilde{H}=0$ in Eq.~(\ref{chiTsol}) yields:
\begin{equation}\label{chiZsolEG}
\chi^{}_z(z)=\sin\left({m\pi\over L}z\right), 
\end{equation}
\begin{equation}\label{chiTsolEG}
\chi^{}_t(t)=1.
\end{equation}
Choosing $C_3=1$ in Eq.~(\ref{chiangpart}) yields (noting that $\upsilon_\theta=D_3=0$ is required for $2\pi$-periodicity):
\begin{equation}\label{chithetaEG}
\chi^{}_\theta(\theta)=1. 
\end{equation}
Since $\upsilon_\theta=0$ and $\upsilon_r>0$, it then follows from Eq.~(\ref{chiradsolNPAP}) that
\begin{equation}\label{chisolcylEG}
\chi^{}_r(r)=A_3I_0\left({m\pi\over L}r\right), 
\end{equation}
where the $K_0({m\pi\over L}r)$ term has been discarded (by prescribing $B_3=0$ in Eq.~(\ref{chiradsolNPAP})) in order to again satisfy the finiteness condition. 

Inserting results (\ref{phipsiZEG})-(\ref{phirssolcyl}) and (\ref{chiZsolEG})-(\ref{chisolcylEG}) into (\ref{DISCOMPeg}) then yields the (general) displacement components:
\begin{subequations}\label{DISCOMPegfin}
\begin{eqnarray}
u_r(r,z,t)&=&-A_2\alpha J_1(\alpha r)\sin\left({k\pi\over L}z\right)\sin(\omega t), \label{RcompEG} \\
u_\theta(r,z,t)&=&A_3I_0\left({m\pi\over L}r\right)\sin\left({m\pi\over L}z\right), \label{THcompEG} \\
u_z(r,z,t)&=&\left({k\pi\over L}\right)\left[\gamma_1A_1+\gamma_2A_2J_0(\alpha r)\right]\cos\left({k\pi\over L}z\right)\sin(\omega t), \label{ZcompEG}
\end{eqnarray}
where $\alpha\equiv\sqrt{|\Lambda_2|}$, and the constants $\gamma_s$ ($s=1,2$) given by Eq.~(\ref{solpsiPP2}) reduce (in this special case) to:
\begin{eqnarray}\label{solpsiPP2EGSS}
\gamma_s=\left \{ \begin{array}{lr}
             1 & ~~\text{if}~s=1 \\
             1-{(\lambda+2\mu)\over\mu} & ~~\text{if}~s=2
           \end{array} \right..
\end{eqnarray}
\end{subequations} 

To complete the solution, we need only to determine the values of the constants $\{A_1,A_2,A_3\}$ in Eqs.~(\ref{RcompEG})-(\ref{ZcompEG}) that satisfy boundary conditions (\ref{BCsCRVD1}). Substituting Eqs.~(\ref{DISCOMPegfin}) into the stress-displacement relations [Eqs.~(\ref{strssdispCYL}) in Appendix \ref{StrssDisRels}] yields (after some algebra) the required stress components as follows\footnote{Although not shown here, it can be verified that the stress component $\sigma_{zz}(r,z,t)$ automatically satisfies boundary condition (\ref{SSBCs3}).}:
\begin{subequations}\label{strsscompEGgen}
\begin{equation}\label{strsscompEGgenRR}
\sigma_{rr}(r,z,t)=-\Bigg\{A_1\lambda\left({k\pi\over L}\right)^2+2A_2\mu\alpha\bigg[\alpha J_0(\alpha r)-{1\over r}J_1(\alpha r)\bigg]\Bigg\}\sin\left({k\pi\over L}z\right)\sin(\omega t),
\end{equation}
\begin{equation}\label{strsscompEGgenRTH}
\sigma_{r\theta}(r,z,t)=A_3\mu\left[\left({m\pi\over L}\right)I_1\left({m\pi\over L}r\right)-{1\over r}I_0\left({m\pi\over L}r\right)\right]\sin\left({m\pi\over L}z\right),
\end{equation}
\begin{equation}\label{strsscompEGgenRZ}
\sigma_{rz}(r,z,t)=A_2\lambda\left({k\pi\over L}\right)\alpha J_1(\alpha r)\cos\left({k\pi\over L}z\right)\sin(\omega t).
\end{equation}
\end{subequations}
Application of boundary conditions (\ref{BCsCRVD1}) then yields the following $3\times3$ linear system:
\begin{eqnarray}\label{BCmatrixEQ2}
\left[\begin{array}{ccc}
            -\lambda \left({k\pi\over L}\right)^2 & -2\mu\alpha J_1'(\alpha R) & 0 \\ 
             0 & 0 & ~~\mu\left[\left({m\pi\over L}\right)I_1\left({m\pi\over L}R\right)-{1\over R}I_0\left({m\pi\over L}R\right)\right] \\
             0 & ~~\lambda\left({k\pi\over L}\right)\alpha J_1(\alpha R) & 0 
           \end{array} \right] \left[\begin{array}{c}
             A_1 \\
             A_2 \\
             A_3 
           \end{array} \right] =  \left[\begin{array}{c}
              \mathcal{A} \\
              \mathcal{B} \\
              \mathcal{C}
           \end{array} \right],
\end{eqnarray}
where $J_1'(\alpha R)\equiv\left[\alpha J_0(\alpha R)-{1\over R}J_1(\alpha R)\right]$. A unique solution to system (\ref{BCmatrixEQ2}) exists only when the following three conditions are satisfied: 
\begin{equation}\label{condsEGsolcyl}
\lambda\neq0, \quad J_1(\alpha R)\neq0, \quad \left({m\pi\over L}R\right)I_1\left({m\pi\over L}R\right)\neq I_0\left({m\pi\over L}R\right). 
\end{equation}
Otherwise the system is underdetermined and there exist an infinite number of solutions. Assuming conditions (\ref{condsEGsolcyl}) are met, system (\ref{BCmatrixEQ2}) can be readily solved yielding  the values:
\begin{subequations}\label{solnconstsolcyl}
\begin{eqnarray}
A_1&=&-{1\over\lambda\left({k\pi\over L}\right)^2}\left(\mathcal{A}+{2\mu J_1'(\alpha R)\over\lambda\left({k\pi\over L}\right)J_1(\alpha R)}\mathcal{C}\right), \\
A_2&=&{\mathcal{C}\over\lambda\left({k\pi\over L}\right)\alpha J_1(\alpha R)}, \\
A_3&=&{\mathcal{B}\over\mu\left[\left({m\pi\over L}\right)I_1\left({m\pi\over L}R\right)-{1\over R}I_0\left({m\pi\over L}R\right)\right]}. 
\end{eqnarray}
\end{subequations} 
The displacement field is therefore given by (\ref{DISCOMPegfin}) with the constants $\{A_1,A_2,A_3\}$ given by (\ref{solnconstsolcyl}). It is worthwhile to reiterate that when all three stress constants $\{\mathcal{A},\mathcal{B},\mathcal{C}\}$ are non-zero, the motion is neither torsional nor axisymmetric; the motion is axisymmetric only when $\mathcal{B}=0$. 

\section{Application 2: Forced Vibration of an Open Thick Cylindrical Shell}

Consider an (elastic) open circular cylindrical shell whose domain is given by: $R_1 \le r \le R_2$, $0\le\theta_1\le\theta\le\theta_2<2\pi$, and $0 \le z \le L$. The geometrical parameters $\{L,R_1,R_2\}$ are all finite, but otherwise arbitrary. The flat ends of the shell, which are situated at $z=0$ and $z=L$, are clamped:  
\begin{subequations}\label{CLMPBCs}
\begin{equation}\label{CLMPBCs1}
u_r(r,\theta,0,t)=u_r(r,\theta,L,t)=0,
\end{equation}
\begin{equation}\label{CLMPBCs2}
u_\theta(r,\theta,0,t)=u_\theta(r,\theta,L,t)=0,
\end{equation}
\begin{equation}\label{CLMPBCs3}
u_{z}(r,\theta,0,t)=u_{z}(r,\theta,L,t)=0. 
\end{equation}
\end{subequations}
Conditions (\ref{CLMPBCs1})-(\ref{CLMPBCs3}) are pure displacement boundary conditions and must be satisfied for all $r\in[R_1,R_2]$, $\theta\in[\theta_1,\theta_2]$, and arbitrary $t$. 

In the following subsections, we shall specify several different sets of boundary conditions on the radial and circumferential faces of the shell and then obtain solutions to the corresponding boundary-value problems. 

\subsection{Linear Circumferential Variations}\label{LinearEG}

Suppose the (radial) curved surfaces of the shell are subjected to the following non-uniform stresses: 
\begin{subequations}\label{BCCURVLN}
\begin{eqnarray}
\sigma_{rr}(R_i,\theta,z,t)&=&\left(\mathcal{A}_i+\mathcal{B}_i\theta\right)\sin\left({k\pi\over L}z\right)\sin(\omega t), \label{LNstrssrr2} \\ 
\sigma_{r\theta}(R_{i},\theta,z,t)&=&\mathcal{C}_{i}\sin\left({k\pi\over L}z\right)\sin(\omega t), \label{LNstrssrt2} \\ 
\sigma_{rz}(R_i,\theta,z,t)&=&\left(\mathcal{F}_i+\mathcal{G}_i\theta\right)\cos\left({k\pi\over L}z\right)\sin(\omega t), \label{LNstrssrz2}
\end{eqnarray}
\end{subequations}
where the index $i=\{1,2\}$, $\{\mathcal{A}_i,\mathcal{B}_i,\mathcal{C}_i,\mathcal{F}_i,\mathcal{G}_i:i=1,2\}$ are prescribed constant stresses, $k\in\mathbb{Z}^+$ is a prescribed constant, and $\omega$ is the prescribed angular excitation frequency. Conditions (\ref{LNstrssrr2})-(\ref{LNstrssrz2}) are pure stress boundary conditions and must be satisfied for all $\theta\in[\theta_1,\theta_2]$, $z\in(0,L)$, and arbitrary $t$. 
Suppose furthermore that the following mixed (non-uniform) conditions are prescribed on the circumferential faces of the shell: 
\begin{subequations}\label{BCTHFACESLN}
\begin{eqnarray}
u_{r}(r,\theta_i,z,t)&=&\mathcal{U}_i\hspace*{0.03cm}R\left({1\over r}\right)\sin\left({k\pi\over L}z\right)\sin(\omega t), \label{LNFstrssrr2} \\ 
u_{z}(r,\theta_i,z,t)&=&0, \label{LNFstrssrz2} \\
\sigma_{\theta\theta}(r,\theta_{i},z,t)&=&\mathcal{S}_iR^2\left({1\over r}\right)^{\hspace*{-0.1cm}2}\sin\left({k\pi\over L}z\right)\sin(\omega t), \label{LNFstrssrt2} 
\end{eqnarray}
\end{subequations}
where again $i=\{1,2\}$, $\{\mathcal{U}_i:i=1,2\}$ are prescribed constant displacements ($\mathcal{U}_1\neq\mathcal{U}_2$), $\{\mathcal{S}_i:i=1,2\}$ are prescribed constant stresses ($\mathcal{S}_1\neq\mathcal{S}_2$), and $R\equiv(R_1+R_2)/2$ is the mean radius of the shell. Conditions (\ref{LNFstrssrr2})-(\ref{LNFstrssrt2}) are admissible mixed boundary conditions specifying orthogonal components of the stress and displacement on the circumferential faces of the shell and must be satisfied for all $r\in[R_1,R_2]$, $z\in(0,L)$, and arbitrary $t$. 

\begin{framed}
\noindent {\textbf{Problem A}: In the special case where the excitation frequency $\omega=\sqrt{\mu\over\rho}\left({k\pi\over L}\right)$, determine the displacement field at all points of the clamped shell subject to conditions (\ref{BCCURVLN}) on its radial surfaces and to conditions (\ref{BCTHFACESLN}) on its circumferential faces.} 
\end{framed}

To obtain general solutions for $\{u_r,u_\theta,u_z\}$ compatible with the given boundary conditions, fix the parameters $\kappa$, $\tau$, and $\eta$ (associated with the Buchwald potentials $\Phi$ and $\Psi$) as in (\ref{kapandtau}) and employ prescription (\ref{sepconsts}) for Buchwald potential $\chi$.   
Then, choosing $E=0$ and $F=1$ in Eq.~(\ref{Zpart}), $G=0$ and $H=1$ in Eq.~(\ref{Tpart}), $\widetilde{E}=0$ and $\widetilde{F}=1$ in Eq.~(\ref{chiZsolMN}), and $\widetilde{G}=0$ and $\widetilde{H}=1$ in Eq.~(\ref{chiTsolMN}) yields:
\begin{equation}\label{phipsiZEG2}
\phi_z(z)=\psi_z(z)=\chi^{}_z(z)=\sin\left({k\pi\over L}z\right), 
\end{equation}
\begin{equation}\label{phipsiTEG2}
\phi_t(t)=\psi_t(t)=\chi^{}_t(t)=\sin(\omega t).
\end{equation}
Since $\eta=0$ [c.f., (\ref{kapandtau})], it follows from Eqs.~(\ref{phiangpartNP}) and (\ref{chiangpartMN}) that: 
\begin{equation}\label{thetafuncsEG2}
\phi^{(1)}_\theta(\theta)=C_1 + D_1\theta, \quad \phi^{(2)}_\theta(\theta)=C_2 + D_2\theta, \quad \chi^{}_\theta(\theta)=C_3 + D_3\theta. 
\end{equation}
The constants $\Lambda_1$ and $\Lambda_2$, as determined from Eq.~(\ref{rootsSC}), are:
\begin{equation}\label{LambdasEG2}
\Lambda_1=\left({k\pi\over L}\right)^2-{\rho\omega^2\over(\lambda+2\mu)}>0, \quad \Lambda_2=\left({k\pi\over L}\right)^2-{\rho\omega^2\over\mu}=0.
\end{equation}
Since $\eta=0$, $\Lambda_1>0$, and $\Lambda_2=0$, it then follows from Eqs.~(\ref{phiradsolNP}) and (\ref{chiradsolNPAPMN}) that:
\begin{equation}\label{phirsEG2}
\phi_r^{(1)}(r)=A_1I_0\left(\sqrt{\Lambda_1}r\right)+B_1K_0\left(\sqrt{\Lambda_1}r\right), ~~ \phi_r^{(2)}(r)=A_2+B_2\ln r, ~~ \chi^{}_r(r)=A_3+B_3\ln r. ~~
\end{equation}
Furthermore, the constants $\gamma_s$ ($s=1,2$) given by Eq.~(\ref{solpsiPP2}) reduce (in this special case) to:
\begin{eqnarray}\label{solpsiPP2EG2}
\gamma_s=\left \{ \begin{array}{lr}
             1 & ~~\text{if}~s=1 \\
             0 & ~~\text{if}~s=2
           \end{array} \right..
\end{eqnarray}
Inputting ingredients (\ref{phipsiZEG2})-(\ref{solpsiPP2EG2}) into (\ref{DISCOMPN2PPkapdiffzero}) and setting $A_1=B_1=0$ in the result (since there are no Bessel functions in the radial parts of (\ref{BCTHFACESLN})) yields the following general displacement field:
\begin{subequations}\label{DisFldGenEG2}
\begin{eqnarray}
u_{r}(r,\theta,z,t)&=&\left[{B_2\over r}\left(C_2+D_2\theta\right)+{D_3\over r}\left(A_3+B_3\ln r\right)\right]\sin\left({k\pi\over L}z\right)\sin(\omega t), \label{DisFldGenEG2A} \\ 
u_{\theta}(r,\theta,z,t)&=&\left[{D_2\over r}\left(A_2+B_2\ln r\right)-{B_3\over r}\left(C_3+D_3\theta\right)\right]\sin\left({k\pi\over L}z\right)\sin(\omega t), \label{DisFldGenEG2B} \\ 
u_{z}(r,\theta,z,t)&=&0. \label{DisFldGenEG2C}
\end{eqnarray}
\end{subequations}
Since there is no ${\ln r\over r}$ term in (\ref{LNFstrssrr2}), we choose $B_3=0$ in (\ref{DisFldGenEG2A}), which consequently eliminates the linear circumferential term in (\ref{DisFldGenEG2B}). Furthermore, since $\mathcal{U}_1\neq\mathcal{U}_2$ in (\ref{LNFstrssrr2}), the constant $B_2$ in (\ref{DisFldGenEG2A}) must be non-zero. By defining a new set of arbitrary constants $\bar{A}_2\equiv A_2D_2$, $\bar{C}_2\equiv B_2C_2+A_3D_3$, and $\bar{D}_2\equiv B_2D_2$, (\ref{DisFldGenEG2}) may then be compactly expressed as follows: 
\begin{subequations}\label{DisFldGenEG2FG}
\begin{eqnarray}
u_{r}(r,\theta,z,t)&=&{1\over r}\left(\bar{C}_2+\bar{D}_2\theta\right)\sin\left({k\pi\over L}z\right)\sin(\omega t), \label{DisFldGenEG2AA} \\ 
u_{\theta}(r,\theta,z,t)&=&{1\over r}\left(\bar{A}_2+\bar{D}_2{\ln r}\right)\sin\left({k\pi\over L}z\right)\sin(\omega t), \label{DisFldGenEG2BA} \\ 
u_{z}(r,\theta,z,t)&=&0. \label{DisFldGenEG2CA}
\end{eqnarray}
\end{subequations}
Note that each of the constants $\{\bar{A}_2,\bar{C}_2,\bar{D}_2\}$ has units of $(\text{length})^2$. Application of boundary condition (\ref{LNFstrssrr2}) using (\ref{DisFldGenEG2AA}) then yields:
\begin{equation}\label{linthetaconsts}
\bar{C}_2=\left({\mathcal{U}_1\theta_2-\mathcal{U}_2\theta_1\over\theta_2-\theta_1}\right)R, \quad \bar{D}_2=\left({\mathcal{U}_2-\mathcal{U}_1\over\theta_2-\theta_1}\right)R.
\end{equation}
Condition (\ref{LNFstrssrz2}) is automatically satisfied by (\ref{DisFldGenEG2CA}). Substituting Eqs.~(\ref{DisFldGenEG2FG}) into the stress-displacement relations [Eqs.~(\ref{strssdispCYL}) in Appendix \ref{StrssDisRels}] yields the pertinent components of the stress field as follows: 
\begin{subequations}\label{strsscompEG2gen}
\begin{eqnarray}
\sigma_{rr}(r,\theta,z,t)&=&-{2\mu\over r^2}\left(\bar{C}_2+\bar{D}_2\theta\right)\sin\left({k\pi\over L}z\right)\sin(\omega t), \label{strsscompEG2genRR} \\
\sigma_{r\theta}(r,\theta,z,t)&=&-{2\mu\over r^2}\left[\left(\bar{A}_2-\bar{D}_2\right)+\bar{D}_2{\ln r}\right]\sin\left({k\pi\over L}z\right)\sin(\omega t), \label{strsscompEG2genRTH} \\ 
\sigma_{rz}(r,\theta,z,t)&=&\left({k\pi\over L}\right){\mu\over r}\left(\bar{C}_2+\bar{D}_2\theta\right)\cos\left({k\pi\over L}z\right)\sin(\omega t), \label{strsscompEG2genRZ} \\ 
\sigma_{\theta\theta}(r,\theta,z,t)&=&{2\mu\over r^2}\left(\bar{C}_2+\bar{D}_2\theta\right)\sin\left({k\pi\over L}z\right)\sin(\omega t). \label{strsscompEG2genTHTH} 
\end{eqnarray}
\end{subequations}
Direct application of boundary conditions (\ref{LNstrssrr2})-(\ref{LNstrssrz2}) and (\ref{LNFstrssrt2}) then yields a conditional solution to Problem A as follows. The displacement field (\ref{DisFldGenEG2FG}) with constants $\bar{C}_2$ and $\bar{D}_2$ given by (\ref{linthetaconsts}) and $\bar{A}_2=\bar{D}_2$ is a solution to Problem A provided the constant stresses $\{\mathcal{A}_i,\mathcal{B}_i,\mathcal{C}_i,\mathcal{F}_i,\mathcal{G}_i,\mathcal{S}_i:i=1,2\}$ are chosen as follows: 
\begin{equation}\label{stressconds}
\begin{split}
\mathcal{A}_i=-{2\mu\over{R}^2_i}\bar{C}_2, \quad \mathcal{B}_i=-{2\mu\over{R}^2_i}\bar{D}_2, \quad \mathcal{C}_i=-2\mu{\ln R_i\over{R}^2_i}\bar{D}_2, \quad \quad \quad \\
\mathcal{F}_i={\mu\over R_i}\left({k\pi\over L}\right)\bar{C}_2, \quad \mathcal{G}_i={\mu\over R_i}\left({k\pi\over L}\right)\bar{D}_2, \quad \mathcal{S}_i={2\mu\over R^2}\left(\bar{C}_2+\bar{D}_2\theta_i\right),  
\end{split}
\end{equation}
where $\bar{C}_2$ and $\bar{D}_2$ are again given by (\ref{linthetaconsts}). 

\subsection{Exponential Circumferential Variations}\label{ExponenEG}

Suppose instead that the radial surfaces of the shell are subjected to the following non-uniform stresses: 
\begin{subequations}\label{BCCURVEXP}
\begin{eqnarray}
\sigma_{rr}(R_i,\theta,z,t)&=&\mathcal{A}_i\exp(-\beta\theta)\sin\left({k\pi\over L}z\right)\sin(\omega t), \label{EXPstrssrr2} \\ 
\sigma_{r\theta}(R_{i},\theta,z,t)&=&\mathcal{B}_{i}\exp(-\beta\theta)\sin\left({k\pi\over L}z\right)\sin(\omega t), \label{EXPstrssrt2} \\ 
\sigma_{rz}(R_i,\theta,z,t)&=&\mathcal{C}_i\exp(-\beta\theta)\cos\left({k\pi\over L}z\right)\sin(\omega t), \label{EXPstrssrz2}
\end{eqnarray}
\end{subequations}
where the index $i=\{1,2\}$, $\{\mathcal{A}_i,\mathcal{B}_i,\mathcal{C}_i:i=1,2\}$ are prescribed constant stresses, $k\in\mathbb{Z}^+$ and $\beta>0$ are prescribed constants, and $\omega$ is the prescribed angular excitation frequency. Conditions (\ref{EXPstrssrr2})-(\ref{EXPstrssrz2}) are pure stress boundary conditions and must be satisfied for all $\theta\in[\theta_1,\theta_2]$, $z\in(0,L)$, and arbitrary $t$. 
Suppose furthermore that the following (non-uniform) conditions are prescribed on the circumferential faces of the shell: 
\begin{subequations}\label{BCTHFACESEXP}
\begin{eqnarray}
u_{r}(r,\theta_i,z,t)&=&\mathcal{D}_i\hspace*{0.03cm}R~{\sin(\beta\ln r)\over r}\sin\left({k\pi\over L}z\right)\sin(\omega t), \label{EXPFstrssrr2} \\ 
u_{\theta}(r,\theta_i,z,t)&=&\mathcal{D}_i\hspace*{0.03cm}R~{\cos(\beta\ln r)\over r}\sin\left({k\pi\over L}z\right)\sin(\omega t), \label{EXPFstrssrt2} \\ 
u_{z}(r,\theta_i,z,t)&=&0, \label{EXPFstrssrz2}  
\end{eqnarray}
\end{subequations}
where again $i=\{1,2\}$, $\{\mathcal{D}_i:i=1,2\}$ are prescribed constant displacements such that $\mathcal{D}_2=\mathcal{D}_1e^{-\beta(\theta_2-\theta_1)}$ and $R\equiv(R_1+R_2)/2$ is the mean radius of the shell. Conditions (\ref{EXPFstrssrr2})-(\ref{EXPFstrssrz2}) are pure displacement boundary conditions on the circumferential faces of the shell and must be satisfied for all $r\in[R_1,R_2]$, $z\in(0,L)$, and arbitrary $t$. 

\begin{framed}
\noindent {\textbf{Problem B}: In the special case where the excitation frequency $\omega=\sqrt{\mu\over\rho}\left({k\pi\over L}\right)$, determine the displacement field at all points of the clamped shell subject to conditions (\ref{BCCURVEXP}) on its radial surfaces and to conditions (\ref{BCTHFACESEXP}) on its circumferential faces.} 
\end{framed}

To obtain general solutions for $\{u_r,u_\theta,u_z\}$ compatible with the given boundary conditions, fix the parameters $\kappa$ and $\tau$ as in (\ref{kapandtau}), fix $\eta=-\beta^2$, and employ prescription (\ref{sepconsts}) for Buchwald potential $\chi$. The axial and temporal parts of the Buchwald potentials are the same as those obtained for Problem A [c.f., Eqs.~(\ref{phipsiZEG2}) and (\ref{phipsiTEG2})]. Since $\eta<0$, it follows from Eqs.~(\ref{phiangpartNP}) and (\ref{chiangpartMN}) that: 
\begin{equation}\label{thetafuncsEG3}
\phi^{(1)}_\theta(\theta)=C_1e^{-\beta\theta} + D_1e^{+\beta\theta}, \quad \phi^{(2)}_\theta(\theta)=C_2e^{-\beta\theta} + D_2e^{+\beta\theta}, \quad \chi^{}_\theta(\theta)=C_3e^{-\beta\theta} + D_3e^{+\beta\theta}. 
\end{equation}
The constants $\Lambda_1$ and $\Lambda_2$ are the same as for Problem A [c.f., Eqs.~(\ref{LambdasEG2})]. 
Since $\eta<0$, $\Lambda_1>0$, and $\Lambda_2=0$, it then follows from Eqs.~(\ref{phiradsolNP}) and (\ref{chiradsolNPAPMN}) that:
\begin{subequations}\label{phirsEG3}
\begin{eqnarray}
\phi_r^{(1)}(r)&=&A_1I_{i\beta}\left(\sqrt{\Lambda_1}r\right)+B_1K_{i\beta}\left(\sqrt{\Lambda_1}r\right), \\ 
\phi_r^{(2)}(r)&=&A_2\cos(\beta\ln r)+B_2\sin(\beta\ln r), \\ 
\chi^{}_r(r)&=&A_3\cos(\beta\ln r)+B_3\sin(\beta\ln r). 
\end{eqnarray}
\end{subequations}
The constants $\gamma_s$ ($s=1,2$) are the same as for Problem A [c.f., Eq.~(\ref{solpsiPP2EG2})]. 
Inputting the above ingredients into (\ref{DISCOMPN2PPkapdiffzero}) and setting $A_1=B_1=0$ in the result (since there are no Bessel functions in the radial parts of (\ref{BCTHFACESEXP})) and furthermore setting $D_1=D_2=D_3=0$ (since there are no $e^{+\beta\theta}$ terms in (\ref{BCCURVEXP})) yields the following general displacement field:
\begin{subequations}\label{DisFldGenEG3}
\begin{eqnarray}
u_{r}(r,\theta,z,t)&=&{\beta\over r}\left\{C_2\left[-A_2\sin(\beta\ln r)+B_2\cos(\beta\ln r)\right]-C_3\left[A_3\cos(\beta\ln r)+B_3\sin(\beta\ln r)\right]\right\} \nonumber \\ 
&&~~~\times\exp({-\beta\theta})\sin\left({k\pi\over L}z\right)\sin(\omega t), \label{DisFldGenEG3A} \\ 
u_{\theta}(r,\theta,z,t)&=&-{\beta\over r}\left\{C_2\left[A_2\cos(\beta\ln r)+B_2\sin(\beta\ln r)\right]+C_3\left[-A_3\sin(\beta\ln r)+B_3\cos(\beta\ln r)\right]\right\} \nonumber \\ 
&&~~~\times\exp({-\beta\theta})\sin\left({k\pi\over L}z\right)\sin(\omega t), \label{DisFldGenEG3B} \\ 
u_{z}(r,\theta,z,t)&=&0. \label{DisFldGenEG3C}
\end{eqnarray}
\end{subequations}
Since there is no $\cos(\beta\ln r)$ term in (\ref{EXPFstrssrr2}) and no $\sin(\beta\ln r)$ term in (\ref{EXPFstrssrt2}), we choose $A_3=B_2=0$ in (\ref{DisFldGenEG3A}) and (\ref{DisFldGenEG3B}). For convenience, and without loss of generality, we may also set $B_3=0$. By defining a new arbitrary constant $\bar{C}\equiv -A_2\beta C_2$, (\ref{DisFldGenEG3}) reduces to: 
\begin{subequations}\label{DisFldGenEG3FG}
\begin{eqnarray}
u_{r}(r,\theta,z,t)&=&\bar{C}{\sin(\beta\ln r)\over r}\exp({-\beta\theta})\sin\left({k\pi\over L}z\right)\sin(\omega t), \label{DisFldGenEG3AA} \\ 
u_{\theta}(r,\theta,z,t)&=&\bar{C}{\cos(\beta\ln r)\over r}\exp({-\beta\theta})\sin\left({k\pi\over L}z\right)\sin(\omega t), \label{DisFldGenEG3BA} \\ 
u_{z}(r,\theta,z,t)&=&0. \label{DisFldGenEG3CA}
\end{eqnarray}
\end{subequations}
Note that the constant $\bar{C}$ has units of $(\text{length})^2$. Application of boundary condition (\ref{EXPFstrssrr2}) using (\ref{DisFldGenEG3AA}) then yields:
\begin{equation}\label{expthetaconst}
\bar{C}=\mathcal{D}_1\exp(\beta\theta_1)R=\mathcal{D}_2\exp(\beta\theta_2)R. 
\end{equation}
Condition (\ref{EXPFstrssrt2}) is then automatically satisfied by (\ref{DisFldGenEG3BA}), and of course condition (\ref{EXPFstrssrz2}) is trivially satisfied by (\ref{DisFldGenEG3CA}). Substituting Eqs.~(\ref{DisFldGenEG3FG}) into the stress-displacement relations [Eqs.~(\ref{strssdispCYL}) in Appendix \ref{StrssDisRels}] then yields the pertinent components of the stress field as follows: 
\begin{subequations}\label{strsscompEG3gen}
\begin{eqnarray}
\sigma_{rr}(r,\theta,z,t)&=&{2\mu\bar{C}\over r^2}\left[\beta\cos(\beta\ln r)-\sin(\beta\ln r)\right]\exp({-\beta\theta})\sin\left({k\pi\over L}z\right)\sin(\omega t), \label{strsscompEG3genRR} \\
\sigma_{r\theta}(r,\theta,z,t)&=&-{2\mu\bar{C}\over r^2}\left[\beta\sin(\beta\ln r)+\cos(\beta\ln r)\right]\exp({-\beta\theta})\sin\left({k\pi\over L}z\right)\sin(\omega t), \label{strsscompEG3genRTH} \\ 
\sigma_{rz}(r,\theta,z,t)&=&\mu\bar{C}\left({k\pi\over L}\right){\sin(\beta\ln r)\over r}\exp({-\beta\theta})\cos\left({k\pi\over L}z\right)\sin(\omega t). \label{strsscompEG3genRZ} 
\end{eqnarray}
\end{subequations}
Direct application of boundary conditions (\ref{EXPstrssrr2})-(\ref{EXPstrssrz2}) then yields a conditional solution to Problem B as follows. The displacement field (\ref{DisFldGenEG3FG}) with constant $\bar{C}$ given by (\ref{expthetaconst}) is a solution to Problem B provided the constant stresses $\{\mathcal{A}_i,\mathcal{B}_i,\mathcal{C}_i:i=1,2\}$ are prescribed as follows: 
\begin{subequations}\label{stresscondsEG3}
\begin{eqnarray}
\mathcal{A}_i&=&{2\mu\over R_i^2}\left[\beta\cos(\beta\ln R_i)-\sin(\beta\ln R_i)\right]\bar{C}, \\ 
\mathcal{B}_i&=&-{2\mu\over R_i^2}\left[\beta\sin(\beta\ln R_i)+\cos(\beta\ln R_i)\right]\bar{C}, \\
\mathcal{C}_i&=&\mu\left({k\pi\over L}\right){\sin(\beta\ln R_i)\over R_i}\bar{C}, 
\end{eqnarray}
\end{subequations}
where $\bar{C}$ is again given by (\ref{expthetaconst}). 

\section{Application 3: Forced Vibration of an Open Solid Cylinder}\label{EGFULL3} 

Consider an (elastic) open circular solid cylinder whose domain is given by: $0 \le r \le R$, $0\le\theta\le{\pi\over\sqrt{101}}$, and $0 \le z \le L$. The flat ends of the cylinder, which are situated at $z=0$ and $z=L$, are supported such that: 
\begin{subequations}\label{MXDBCs}
\begin{equation}\label{MXDBCs1}
\sigma_{rz}(r,\theta,0,t)=\sigma_{rz}(r,\theta,L,t)=0,
\end{equation}
\begin{equation}\label{MXDBCs2}
\sigma_{\theta z}(r,\theta,0,t)=\sigma_{\theta z}(r,\theta,L,t)=0,
\end{equation}
\begin{equation}\label{MXDBCs3}
u_{z}(r,\theta,0,t)=u_{z}(r,\theta,L,t)=0. 
\end{equation}
\end{subequations}
Conditions (\ref{MXDBCs}) are admissible mixed boundary conditions specifying orthogonal components of the stress and displacement at the flat ends of the cylinder and must be satisfied for all $r\in[0,R]$, $\theta\in[0,\pi/\sqrt{101}]$, and arbitrary $t$. The radial curved surface of the cylinder is subjected to the following non-uniform stresses: 
\begin{subequations}\label{BCCURVTRIG}
\begin{eqnarray}
\sigma_{rr}(R,\theta,z,t)&=&\mathcal{A}\sin\left(\sqrt{101}\hspace*{0.05cm}\theta\right)\sin(\omega t), \label{TRIGstrssrr2} \\ 
\sigma_{r\theta}(R,\theta,z,t)&=&\mathcal{B}\cos\left(\sqrt{101}\hspace*{0.05cm}\theta\right)\sin(\omega t), \label{TRIGstrssrt2} \\ 
\sigma_{rz}(R,\theta,z,t)&=&0, \label{TRIGstrssrz2}
\end{eqnarray}
\end{subequations}
where $\{\mathcal{A},\mathcal{B}\}$ are prescribed constant stresses and $\omega$ is the prescribed angular excitation frequency. Conditions (\ref{BCCURVTRIG}) are pure stress boundary conditions and must be satisfied for all $\theta\in[0,\pi/\sqrt{101}]$, $z\in(0,L)$, and arbitrary $t$. 
The following uniform mixed conditions are furthermore prescribed on the circumferential faces of the cylinder: 
\begin{subequations}\label{BCTHFACESTRIG}
\begin{eqnarray}
u_{r}(r,0,z,t)&=&u_{r}(r,{\pi\over\sqrt{101}},z,t)=0, \label{TRIGFstrssrr2} \\ 
\sigma_{\theta\theta}(r,0,z,t)&=&\sigma_{\theta\theta}(r,{\pi\over\sqrt{101}},z,t)=0, \label{TRIGFstrssrt2} \\ 
u_{z}(r,0,z,t)&=&u_{z}(r,{\pi\over\sqrt{101}},z,t)=0. \label{TRIGFstrssrz2} 
\end{eqnarray}
\end{subequations}
Conditions (\ref{BCTHFACESTRIG}) are admissible mixed boundary conditions specifying orthogonal components of the stress and displacement on the circumferential faces of the cylinder and must be satisfied for all $r\in[0,R]$, $z\in(0,L)$, and arbitrary $t$. 

\begin{framed}
\noindent {\textbf{Problem C}: For arbitrary $\omega>0$, determine the displacement field at all points of the cylinder subject to: (i) conditions (\ref{MXDBCs}) at its flat ends; (ii) conditions (\ref{BCCURVTRIG}) on its radial curved surface; and (iii) conditions (\ref{BCTHFACESTRIG}) on its circumferential faces.} 
\end{framed}

The fact that conditions (\ref{BCCURVTRIG}) are independent of $z$ hints that we should employ the $\kappa=0$ solution (\ref{finsolnkapeq0}) with the axial parts of the Buchwald potentials equal to a constant. For convenience, we can choose $\left\{E=1,F=0\right\}$ in Eq.~(\ref{Zpartkapeq0}) and $\left\{\widetilde{E}=1,\widetilde{F}=0\right\}$ in Eq.~(\ref{ZpartCkapeq0}), which yields: 
\begin{equation}\label{phipsiZEG3}
\phi_z(z)=\psi_z(z)=\chi^{}_z(z)=1. 
\end{equation}
The parameter $\tau$ can be set as in the previous two vibration problems (i.e., $\tau=-\omega^2<0$) to generate the time dependence found in conditions (\ref{BCCURVTRIG}). Choosing $\left\{G=0,H=1\right\}$ in Eq.~(\ref{Tpart}) and $\left\{\widetilde{G}=0,\widetilde{H}=1\right\}$ in Eq.~(\ref{chiTsolMN}) then reproduces (\ref{phipsiTEG2}). Solution (\ref{finsolnkapeq0}) subsequently reduces to:
\begin{subequations}\label{finsolnkapeq0EG31}
\begin{eqnarray}
u_r&=&\left[{\text{d}R_1(r)\over\text{d}r}\Theta_1(\theta)+{1\over r}\chi^{}_r(r){\text{d}\chi^{}_\theta(\theta)\over\text{d}\theta}\right]\sin(\omega t), \label{solnradcompkapeq0EG31} \\
u_\theta&=&\left[{1\over r}R_1(r){\text{d}\Theta_1(\theta)\over\text{d}\theta}-{\text{d}\chi^{}_r(r)\over\text{d}r}\chi^{}_\theta(\theta)\right]\sin(\omega t), \label{solnangcompkapeq0EG31} \\
u_z&=&0. \label{solnaxcompkapeq0EG31}
\end{eqnarray}
\end{subequations}
By substituting (\ref{finsolnkapeq0EG31}) into (\ref{stssstrnCYL1}) and comparing the result with (\ref{TRIGstrssrr2}), it can be deduced that the functions $\left\{\Theta_1(\theta),{\text{d}^2\Theta_1(\theta)\over\text{d}\theta^2},{\text{d}\chi^{}_\theta(\theta)\over\text{d}\theta}\right\}$ must each be of the form: $\sin\left(\sqrt{101}\hspace*{0.05cm}\theta\right)$. Setting the parameter $\eta=101>0$, and then choosing $\left\{C_1=0,D_1=1\right\}$ in Eq.~(\ref{solphiPkapeq0B}) and $\left\{C_3=1,D_3=0\right\}$ in Eq.~(\ref{solchiperpkapeq0A}) gives:
\begin{equation}\label{thetapartsEG3}
\Theta_1(\theta)=\sin\left(\sqrt{101}\hspace*{0.05cm}\theta\right), \quad \chi^{}_\theta(\theta)=\cos\left(\sqrt{101}\hspace*{0.05cm}\theta\right).
\end{equation}
Then, choosing $B_1=0$ in Eq.~(\ref{solphiPkapeq0C}) and $B_3=0$ in Eq.~(\ref{solchiperpkapeq0B}) (in order to satisfy the implicit finiteness condition at $r=0$) subsequently yields (since $\tau<0$ and $\eta>0$):
\begin{equation}\label{radpartsEG3}
R_1(r)=A_1J_{\sqrt{101}}(\alpha_1r), \quad \chi^{}_r(r)=A_3J_{\sqrt{101}}(\alpha_2r),
\end{equation}
where $\alpha_1=\sqrt{\rho\omega^2\over\lambda+2\mu}$ and $\alpha_2=\sqrt{\rho\omega^2\over\mu}$. Inputting (\ref{thetapartsEG3}) and (\ref{radpartsEG3}) into (\ref{finsolnkapeq0EG31}) then yields the displacement field: 
\begin{subequations}\label{finsolnkapeq0EG32}
\begin{eqnarray}
u_r&=&\left[A_1J'_{\sqrt{101}}(\alpha_1r)-A_3{\sqrt{101}\over r}J_{\sqrt{101}}(\alpha_2r)\right]\sin\left(\sqrt{101}\hspace*{0.05cm}\theta\right)\sin(\omega t), \label{solnradcompkapeq0EG32} \\ 
u_\theta&=&\left[A_1{\sqrt{101}\over r}J_{\sqrt{101}}(\alpha_1r)-A_3J'_{\sqrt{101}}(\alpha_2r)\right]\cos\left(\sqrt{101}\hspace*{0.05cm}\theta\right)\sin(\omega t), \label{solnangcompkapeq0EG32} \\ 
u_z&=&0,\label{solnaxcompkapeq0EG32} 
\end{eqnarray}
\end{subequations}
where the primes denote differentiation with respect to $r$. Note that displacement field (\ref{finsolnkapeq0EG32}) automatically satisfies conditions (\ref{MXDBCs3}), (\ref{TRIGFstrssrr2}), and (\ref{TRIGFstrssrz2}). 
Substituting (\ref{finsolnkapeq0EG32}) into the stress-displacement relations [Eqs.~(\ref{strssdispCYL}) in Appendix \ref{StrssDisRels}] then yields the pertinent components of the stress field as follows: 
\begin{subequations}\label{strsscompEG33gen}
\begin{eqnarray}
\sigma_{rr}(r,\theta,z,t)&=&\left\{A_1\left[(\lambda+2\mu)J''_{\sqrt{101}}(\alpha_1r)+{\lambda\over r}J'_{\sqrt{101}}(\alpha_1r)-{101\lambda\over r^2}J_{\sqrt{101}}(\alpha_1r)\right]\right. \nonumber \\ 
&+&\left.A_3{2\mu\sqrt{101}\over r}\left[{1\over r}J_{\sqrt{101}}(\alpha_2r)-J'_{\sqrt{101}}(\alpha_2r)\right]\right\}\sin\left(\sqrt{101}\hspace*{0.05cm}\theta\right)\sin(\omega t), \label{strsscompEG33genRR} \\
\sigma_{r\theta}(r,\theta,z,t)&=&\mu\left\{A_3\left[-J''_{\sqrt{101}}(\alpha_2r)+{1\over r}J'_{\sqrt{101}}(\alpha_2r)-{101\over r^2}J_{\sqrt{101}}(\alpha_2r)\right]\right. \nonumber \\ 
&+&\left.A_1{2\sqrt{101}\over r}\left[J'_{\sqrt{101}}(\alpha_1r)-{1\over r}J_{\sqrt{101}}(\alpha_1r)\right]\right\}\cos\left(\sqrt{101}\hspace*{0.05cm}\theta\right)\sin(\omega t), \label{strsscompEG33genRTH} \\ 
\sigma_{\theta\theta}(r,\theta,z,t)&=&\left\{A_1\left[\lambda J''_{\sqrt{101}}(\alpha_1r)+{(\lambda+2\mu)\over r}\left(J'_{\sqrt{101}}(\alpha_1r)-{101\over r}J_{\sqrt{101}}(\alpha_1r)\right)\right]\right. \nonumber \\ 
&+&\left.A_3{2\mu\sqrt{101}\over r}\left[J'_{\sqrt{101}}(\alpha_2r)-{1\over r}J_{\sqrt{101}}(\alpha_2r)\right]\right\}\sin\left(\sqrt{101}\hspace*{0.05cm}\theta\right)\sin(\omega t),  \label{strsscompEG33genTHTH} \\ 
\sigma_{rz}(r,\theta,z,t)&=&0, \label{strsscompEG33genRZ} \\ 
\sigma_{\theta z}(r,\theta,z,t)&=&0, \label{strsscompEG33genTHZED}
\end{eqnarray}
\end{subequations}
where the primes again denote differentiation with respect to $r$. 

By virtue of (\ref{strsscompEG33genTHTH}) and (\ref{strsscompEG33genTHZED}), conditions (\ref{TRIGFstrssrt2}) and (\ref{MXDBCs2}) are respectively satisfied. And by virtue of (\ref{strsscompEG33genRZ}), conditions (\ref{MXDBCs1}) and (\ref{TRIGstrssrz2}) are naturally satisfied. Thus, the only conditions that remain to be satisfied are conditions (\ref{TRIGstrssrr2}) and (\ref{TRIGstrssrt2}). Application of these two remaining boundary conditions yields the following $2\times2$ linear system:
\begin{subequations}\label{BCmatrixEQEG3}
\begin{eqnarray}
\left[\begin{array}{cc}
             a_{11} & a_{12}  \\
             a_{21} & a_{22} 
           \end{array} \right] \left[\begin{array}{c}
             A_1 \\
             A_3
           \end{array} \right] =  \left[\begin{array}{c}
              \mathcal{A} \\
              \mathcal{B}
           \end{array} \right], 
\end{eqnarray} 
where\footnote{Although the matrix elements (\ref{EG3ME1})-(\ref{EG3ME4}) can be written in several different forms free of derivatives, it is not relevant for our purposes here to do so.} 
\begin{eqnarray}
a_{11}&=&(\lambda+2\mu)J''_{\sqrt{101}}(\alpha_1R)+{\lambda\over R}J'_{\sqrt{101}}(\alpha_1R)-{101\lambda\over R^2}J_{\sqrt{101}}(\alpha_1R), \label{EG3ME1} \\ 
a_{12}&=&{2\mu\sqrt{101}\over R}\left[{1\over R}J_{\sqrt{101}}(\alpha_2R)-J'_{\sqrt{101}}(\alpha_2R)\right], \label{EG3ME2} \\
a_{21}&=&{2\mu\sqrt{101}\over R}\left[J'_{\sqrt{101}}(\alpha_1R)-{1\over R}J_{\sqrt{101}}(\alpha_1R)\right], \label{EG3ME3} \\
a_{22}&=&\mu\left[-J''_{\sqrt{101}}(\alpha_2R)+{1\over R}J'_{\sqrt{101}}(\alpha_2R)-{101\over R^2}J_{\sqrt{101}}(\alpha_2R)\right], \label{EG3ME4} 
\end{eqnarray}
and 
\begin{equation}
J'_{\sqrt{101}}(\alpha_{1/2} R)\equiv\left.{\text{d}J_{\sqrt{101}}(\alpha_{1/2} r)\over\text{d}r}\right|_{r=R}, \quad J''_{\sqrt{101}}(\alpha_{1/2} R)\equiv\left.{\text{d}^2J_{\sqrt{101}}(\alpha_{1/2} r)\over\text{d}r^2}\right|_{r=R}.
\end{equation}
\end{subequations}
The solution to system (\ref{BCmatrixEQEG3}) can then be readily obtained as follows:
\begin{equation}\label{solntomatEQEG3}
A_1={a_{22}\mathcal{A}-a_{12}\mathcal{B}\over a_{11}a_{22}-a_{12}a_{21}}, \quad A_3={a_{11}\mathcal{B}-a_{21}\mathcal{A}\over a_{11}a_{22}-a_{12}a_{21}},
\end{equation}
where $\{a_{11},a_{12},a_{21},a_{22}\}$ are as given by Eqs.~(\ref{EG3ME1})-(\ref{EG3ME4}). The solution to Problem C is therefore given by the displacement field (\ref{finsolnkapeq0EG32}) with the constants $\left\{A_1,A_3\right\}$ given by (\ref{solntomatEQEG3}). 


\section{Conclusion}

In this paper, we have extended the work of SC1 to encompass a much broader class of separable solutions to the coupled equations of motion governing the Buchwald displacement potentials in cylindrical coordinates, namely, separable solutions whose circumferential parts are elementary $2\pi$-aperiodic functions. Using the Buchwald representation (\ref{Buchy}), we then constructed eighteen distinct families of separable solutions to the NL equation; in each case, the circumferential part of the solution is one of three elementary $2\pi$-aperiodic functions. To our knowledge, these solutions constitute the first comprehensive set of $2\pi$-aperiodic cylindrical solutions to the NL equation. The solutions obtained in SC1 can reproduced by taking appropriate values of the angular parameter $\eta$ and are thus special cases of the solutions obtained here. 

The obtained solutions can be directly applied to a fundamental set of linear-elastic boundary-value problems in cylindrical coordinates, in particular, problems wherein the displacement, stress, and strain fields are: constant, linear, sinusoidal, exponential, or hyperbolic in the circumferential and longitudinal coordinates, and sinusoidal or exponential in the time coordinate. Such solutions are useful for solving a wide variety of dynamical problems where $2\pi$-periodicity in the angular coordinate is incompatible with the given boundary conditions. As illustrative examples, we showed how the obtained solutions can be used to solve various forced-vibration problems involving open thick cylindrical shells and open solid cylinders. As an addendum to SC1, we also included an example that illustrates the use of the special $2\pi$-periodic solution (\ref{DISCOMPeg}), an important subsolution that is useful for problems where there is no explicit dependence on the circumferential coordinate.

The applications considered in this paper serve as a good starting point insofar as they are mathematically non-trivial but not overly complicated. More interesting and mathematically complex problems can be solved with the aid of the obtained solutions, but it is not within the scope of the present paper to do so. We hope to report on more advanced applications elsewhere. An important facet that we have not considered here is the rather large number of special solutions that derive from assuming one or more of the Buchwald potentials are zero, which as it turns out can yield more varieties of solutions and thus allow for more interesting applications. We hope to report on this topic in a future paper. 

\section*{Acknowledgments}
The authors acknowledge financial support from the Natural Sciences and Engineering Research Council (NSERC) of Canada and the Ontario Research Foundation (ORF). 

\appendix

\section{Fundamental Separable Solutions to the 2D Helmholtz Equation in Polar Coordinates}\label{twodimHelmPCs}

In polar coordinates, the 2D Helmholtz equation is given by: 
\begin{equation}\label{2DLapEq}
\left(\nabla_\perp^2+\Lambda\right)f=0 \quad \Leftrightarrow \quad {\partial^2f\over\partial r^2} + {1\over r}{\partial f\over\partial r} + {1\over r^2}{\partial^2f\over\partial \theta^2}+\Lambda f=0,
\end{equation}
where $f(r,\theta)$ is some real-valued function of interest and $\Lambda\in\mathbb{R}$ is the Helmholtz constant. Fundamental separable solutions to (\ref{2DLapEq}) can be obtained by assuming a solution of the form $f(r,\theta)=R(r)\Theta(\theta)$ and using separation of variables. Under the assumption that the angular function $\Theta(\theta)$ is $2\pi$-periodic, the solutions are well-known and readily available. The complementary solutions lacking $2\pi$-periodicity do not however appear to be well-known nor are they (to our knowledge) readily found in the literature on PDEs. Since these complementary (i.e., non-$2\pi$-periodic) solutions are fundamental building blocks in our mathematical construction, it is useful to establish them here for referential purposes. 

Substituting $f(r,\theta)=R(r)\Theta(\theta)$ into Eq.~(\ref{2DLapEq}) and then multiplying the result by $r^2/R(r)\Theta(\theta)$ yields (after rearrangement of the radial and angular terms):
\begin{equation}\label{SepEqs}
r^2{R''(r)\over R(r)}+r{R'(r)\over R(r)}+\Lambda r^2=-{\Theta''(\theta)\over\Theta(\theta)}. 
\end{equation}
Since the left-hand side of Eq.~(\ref{SepEqs}) depends only on $r$ and the right-hand side depends only on $\theta$, Eq.~(\ref{SepEqs}) can be satisfied for all values of $r$ and $\theta$ only if the terms on each side of the equation are equal to a constant. If we denote this constant by $\eta$, this leads to the equations:
\begin{equation}\label{HReqn}
r^2R''(r)+rR'(r)+\left(\Lambda r^2-\eta\right)R(r)=0,
\end{equation}
\begin{equation}\label{HTheqn}
\Theta''(\theta)+\eta\Theta(\theta)=0.
\end{equation}
The solution to Eq.~(\ref{HTheqn}) is:
\begin{eqnarray}\label{solntoHTheqn}
\Theta(\theta)=\left \{ \begin{array}{lr}
             C\exp\left(-\sqrt{|\eta|}\theta\right)+D\exp\left(\sqrt{|\eta|}\theta\right) & \text{if}~\eta<0 \\
             C + D\theta & \text{if}~\eta=0 \\
             C\cos\left(\sqrt{\eta}\theta\right)+D\sin\left(\sqrt{\eta}\theta\right) & \text{if}~\eta>0
           \end{array} \right.,
\end{eqnarray}
where $\{C,D\}$ are arbitrary constants. In the following, it will be assumed that $\eta\in\mathbb{R}\backslash\left\{N^2:N\in\mathbb{Z}^+\right\}$. The form of the solution to Eq.~(\ref{HReqn}) depends on the relative signs of $\Lambda$ and $\eta$ as follows. 

\subsection{$\Lambda>0$ and $\eta\ge0$}

In this case, Eq.~(\ref{HReqn}) is a standard (unmodified) Bessel equation with solution: 
\begin{equation}\label{solns1R}
R(r)=AJ_{\sqrt{\eta}}\left(\sqrt{\Lambda}r\right)+BY_{\sqrt{\eta}}\left(\sqrt{\Lambda}r\right),
\end{equation}
where $\{A,B\}$ are arbitrary constants. 

\subsection{$\Lambda>0$ and $\eta<0$}

In this case, Eq.~(\ref{HReqn}) can be expressed in the form:
\begin{equation}\label{HReqnF2}
r^2R''(r)+rR'(r)+\left(\Lambda r^2+|\eta|\right)R(r)=0, 
\end{equation}
which is again an unmodified Bessel equation whose solution is given by:
\begin{equation}\label{solns2R}
R(r)=AJ_{i\sqrt{|\eta|}}\left(\sqrt{\Lambda}r\right)+BY_{i\sqrt{|\eta|}}\left(\sqrt{\Lambda}r\right).
\end{equation}
Note that the Bessel functions in Eq.~(\ref{solns2R}) have purely imaginary orders, and are therefore complex-valued (see Ref.~\cite{Dunster90} for more information on such Bessel functions). 

\subsection{$\Lambda<0$ and $\eta\ge0$}

In this case, Eq.~(\ref{HReqn}) is of the form:
\begin{equation}\label{HReqnF3}
r^2R''(r)+rR'(r)-\left(|\Lambda| r^2+\eta\right)R(r)=0, 
\end{equation}
which is a modified Bessel equation with solution:
\begin{equation}\label{solns3R}
R(r)=AI_{\sqrt{\eta}}\left(\sqrt{|\Lambda|}r\right)+BK_{\sqrt{\eta}}\left(\sqrt{|\Lambda|}r\right).
\end{equation}

\subsection{$\Lambda<0$ and $\eta<0$}

In this case, Eq.~(\ref{HReqn}) can be expressed in the form:
\begin{equation}\label{HReqnF4}
r^2R''(r)+rR'(r)-\left(|\Lambda| r^2-|\eta|\right)R(r)=0,
\end{equation}
which is again a modified Bessel equation with solution given by:
\begin{equation}\label{solns4R}
R(r)=AI_{i\sqrt{|\eta|}}\left(\sqrt{|\Lambda|}r\right)+BK_{i\sqrt{|\eta|}}\left(\sqrt{|\Lambda|}r\right).
\end{equation}
Note that the modified Bessel functions in Eq.~(\ref{solns4R}) have purely imaginary orders; in this case, $I_{i\sqrt{|\eta|}}(\cdots)$ is complex-valued, whereas $K_{i\sqrt{|\eta|}}(\cdots)$ is real-valued. 

\subsection{$\Lambda=0$}

In this case, Eq.~(\ref{HReqn}) reduces to:
\begin{equation}\label{HReqnF5}
r^2R''(r)+rR'(r)-\eta R(r)=0,
\end{equation}
which is a Cauchy-Euler (equidimensional) equation whose solution (depending on the value of $\eta$) is given by: 
\begin{eqnarray}\label{solns5R}
R(r)=\left \{ \begin{array}{lr}
              Ar^{\sqrt{\eta}}+Br^{-\sqrt{\eta}} & \text{if}~\eta>0 \\
              A+B\ln r& \text{if}~\eta=0 \\
              A\cos\left(\sqrt{|\eta|}\ln r\right)+B\sin\left(\sqrt{|\eta|}\ln r\right)& \text{if}~\eta<0
           \end{array} \right..
\end{eqnarray}

\section{A Brief on Bessel Functions with Purely Imaginary Orders}\label{APPBSSLIMG}


For $\nu\in\mathbb{R}$ and $x\in(0,\infty)$, the functions $\displaystyle \left\{I_{i\nu}\left(x\right),K_{i\nu}\left(x\right)\right\}$ are linearly independent solutions of
\begin{equation}\label{BsslC4}
x^2{\mathrm{d}^2y\over\mathrm{d}x^2}+x{\mathrm{d}y\over\mathrm{d}x}+\left(\nu^2-x^2\right)y=0.
\end{equation}
The functions $\displaystyle I_{i\nu}\left(x\right)$ and $\displaystyle K_{i\nu}\left(x\right)$ are complex- and real-valued, respectively. It can be shown that the Wronskian of $\displaystyle  \bar{I}_{\nu}\left(x\right) \equiv \text{Re}\left\{I_{i\nu}\left(x\right)\right\}$ and $\displaystyle K_{i\nu}\left(x\right)$ is non-zero \cite{NISTHB}, and thus these two functions together constitute a real and linearly independent set of solutions to Eq.~(\ref{BsslC4}).

For $\nu\in\mathbb{R}$ and $x\in(0,\infty)$, the functions $\displaystyle \left\{J_{i\nu}\left(x\right),Y_{i\nu}\left(x\right)\right\}$ are complex-valued linearly independent solutions of
\begin{equation}\label{BsslC2}
x^2{\mathrm{d}^2y\over\mathrm{d}x^2}+x{\mathrm{d}y\over\mathrm{d}x}+\left(x^2+\nu^2\right)y=0.
\end{equation}
It can be shown that the Wronskian of the real functions
\begin{subequations}
\begin{eqnarray}
\bar{J}_{\nu}\left(x\right)&\equiv&\text{sech}\left({\pi\nu\over2}\right)\text{Re}\left\{J_{i\nu}\left(x\right)\right\} \\
\overline{Y}_{\nu}\left(x\right)&\equiv&\text{sech}\left({\pi\nu\over2}\right)\text{Re}\left\{Y_{i\nu}\left(x\right)\right\}
\end{eqnarray}
\end{subequations}
is non-zero \cite{NISTHB}, and thus these functions together constitute a real and linearly independent set of solutions to Eq.~(\ref{BsslC2}).

\section{Solution to Subsystem (\ref{systemeSFIN3}) using Separation of Variables}\label{AppSOV}

\noindent Substituting (\ref{chiprodsol}) into Eq.~(\ref{systemeSFIN3}) and then dividing the result by $\chi^{}_r(r)\chi^{}_\theta(\theta)\chi^{}_z(z)\chi^{}_t(t)$ yields
\begin{equation}\label{SOVEq}
\left[{\chi_r''(r)\over\chi^{}_r(r)} + {1\over r}{\chi_r'(r)\over\chi^{}_r(r)} + {1\over r^2} {\chi_\theta''(\theta)\over\chi^{}_\theta(\theta)}\right] + \left[{\chi_z''(z)\over\chi^{}_z(z)}\right] = \left[{1\over c_T^2}{\chi_t''(t)\over\chi^{}_t(t)}\right],
\end{equation}
where $\displaystyle c_T\equiv\sqrt{\mu/\rho}$. The first bracketed term depends only on $r$ and $\theta$, the second bracketed term only on $z$, and the right-hand side depends only on $t$. The above equation can be satisfied for all values of $r$, $\theta$, $z$, and $t$ only if each of the bracketed terms is a constant. This leads to the equations
\begin{equation}\label{Teq}
\chi_t''(t)-\upsilon_tc_T^2\chi^{}_t(t)=0,
\end{equation}
\begin{equation}\label{Zeq}
\chi_z''(z)-\upsilon_z\chi^{}_z(z)=0,
\end{equation}
and
\begin{equation}\label{RandTHeq}
\left[{\chi_r''(r)\over\chi^{}_r(r)} + {1\over r}{\chi_r'(r)\over\chi^{}_r(r)} + {1\over r^2} {\chi_\theta''(\theta)\over\chi^{}_\theta(\theta)}\right]=\upsilon_r,
\end{equation}
where the separation constants $\upsilon_t$, $\upsilon_z$, and $\upsilon_r$ obey (by virtue of Eq.~(\ref{SOVEq})) the relation
\begin{equation}\label{constsrel}
\upsilon_r+\upsilon_z=\upsilon_t.
\end{equation}
Equation (\ref{RandTHeq}) can be written in the form
\begin{equation}\label{RandTHeq2}
r^2{\chi_r''(r)\over\chi^{}_r(r)} + r{\chi_r'(r)\over\chi^{}_r(r)} -r^2\upsilon_r = -{\chi_\theta''(\theta)\over\chi^{}_\theta(\theta)}.
\end{equation}
The left-hand side of Eq.~(\ref{RandTHeq2}) depends only on $r$ and the right-hand side only on $\theta$. The two sides must therefore be equal to another constant, which we shall denote by $\upsilon_\theta$. This second separation of variables yields the equations
\begin{equation}\label{DEchiradpart}
r^2\chi_r''(r) +r\chi_r'(r) -\left(r^2\upsilon_r+\upsilon_\theta\right)\chi^{}_r(r)=0,
\end{equation}
\begin{equation}\label{DEchiangpart}
\chi_\theta''(\theta)+\upsilon_\theta\chi^{}_\theta(\theta)=0.
\end{equation}
Note that all real values of the separation constants $\upsilon_\theta$, $\upsilon_t$, $\upsilon_z$, and $\upsilon_r$ are admissible but that the latter three constants are not all independent. Upon independent specification of any two of $\{\upsilon_t,\upsilon_z, \upsilon_r\}$, the third is automatically determined by relation (\ref{constsrel}). Without loss of generality, we may assign $\upsilon_t$ and $\upsilon_z$ to be independent parameters (that can take on any real values) whereupon $\upsilon_r=\upsilon_t-\upsilon_z$. Solutions to Eqs.~(\ref{DEchiangpart}), (\ref{Zeq}), and (\ref{Teq}) are then given by: 
\begin{eqnarray}\label{chiangpart}
\chi^{}_\theta(\theta)=\left \{ \begin{array}{lr}
             C_3\exp\left(-\sqrt{|\upsilon_\theta|}\theta\right)+D_3\exp\left(\sqrt{|\upsilon_\theta|}\theta\right) & \text{if}~\upsilon_\theta<0 \\
             C_3 + D_3\theta & \text{if}~\upsilon_\theta=0 \\
             C_3\cos\left(\sqrt{\upsilon_\theta}\theta\right)+D_3\sin\left(\sqrt{\upsilon_\theta}\theta\right) & \text{if}~\upsilon_\theta>0
           \end{array} \right.,
\end{eqnarray}
\begin{eqnarray}\label{chiZsol}
\chi^{}_z(z)=\left \{ \begin{array}{lr}
             \widetilde{E}\cos\left(\sqrt{|\upsilon_z|}z\right)+\widetilde{F}\sin\left(\sqrt{|\upsilon_z|}z\right) & \text{if}~\upsilon_z<0 \\
             \widetilde{E} + \widetilde{F}z & \text{if}~\upsilon_z=0 \\
             \widetilde{E}\exp\left(-\sqrt{\upsilon_z}z\right)+\widetilde{F}\exp\left(\sqrt{\upsilon_z}z\right) & \text{if}~\upsilon_z>0
           \end{array} \right.,
\end{eqnarray}  
and
\begin{eqnarray}\label{chiTsol}
\chi^{}_t(t)=\left \{ \begin{array}{lr}
             \widetilde{G}\cos\left(\sqrt{|\upsilon_t|}c_Tt\right)+\widetilde{H}\sin\left(\sqrt{|\upsilon_t|}c_Tt\right) & \text{if}~\upsilon_t<0 \\
             \widetilde{G} + \widetilde{H}t & \text{if}~\upsilon_t=0 \\
             \widetilde{G}\exp\left(-\sqrt{\upsilon_t}c_Tt\right)+\widetilde{H}\exp\left(\sqrt{\upsilon_t}c_Tt\right) & \text{if}~\upsilon_t>0
           \end{array} \right.,
\end{eqnarray}  
respectively, where $\displaystyle\left\{C_3,D_3,\widetilde{E},\widetilde{F},\widetilde{G},\widetilde{H}\right\}$ are arbitrary constants. The function $\chi^{}_\theta(\theta)$ given by Eq.~(\ref{chiangpart}) is $2\pi$-periodic when $\upsilon_\theta\in\left\{N^2:N\in\mathbb{Z}^+\right\}$ and in the special case $\upsilon_\theta=D_3=0$. 
Using the results of Appendix \ref{twodimHelmPCs}, solutions to Eq.~(\ref{DEchiradpart}) can be immediately deduced; the solutions are as follows: 
\begin{eqnarray}\label{chiradsolNPAP}
\chi^{}_r(r)=\left \{ \begin{array}{lr}
           A_3J_{i\sqrt{|\upsilon_\theta|}}\left(\sqrt{|\upsilon_r|}r\right)+B_3Y_{i\sqrt{|\upsilon_\theta|}}\left(\sqrt{|\upsilon_r|}r\right) & \text{if}~\upsilon_r<0~\text{and}~\upsilon_\theta<0 \\
            A_3\cos\left(\sqrt{|\upsilon_\theta|}\ln r\right)+B_3\sin\left(\sqrt{|\upsilon_\theta|}\ln r\right) & \text{if}~\upsilon_r=0~\text{and}~\upsilon_\theta<0 \\
           A_3I_{i\sqrt{|\upsilon_\theta|}}\left(\sqrt{\upsilon_r}r\right)+B_3K_{i\sqrt{|\upsilon_\theta|}}\left(\sqrt{\upsilon_r}r\right) & \text{if}~\upsilon_r>0~\text{and}~\upsilon_\theta<0 \\
           A_3J_0\left(\sqrt{|\upsilon_r|}r\right)+B_3Y_0\left(\sqrt{|\upsilon_r|}r\right) & \text{if}~\upsilon_r<0~\text{and}~\upsilon_\theta=0 \\
           A_3+B_3\ln r & \text{if}~\upsilon_r=0~\text{and}~\upsilon_\theta=0 \\ 
           A_3I_0\left(\sqrt{\upsilon_r}r\right)+B_3K_0\left(\sqrt{\upsilon_r}r\right) & \text{if}~\upsilon_r>0~\text{and}~\upsilon_\theta=0 \\
           A_3J_{\sqrt{\upsilon_\theta}}\left(\sqrt{|\upsilon_r|}r\right)+B_3Y_{\sqrt{\upsilon_\theta}}\left(\sqrt{|\upsilon_r|}r\right) & \text{if}~\upsilon_r<0~\text{and}~\upsilon_\theta>0 \\ 
            A_3r^{\sqrt{\upsilon_\theta}}+B_3r^{-\sqrt{\upsilon_\theta}} & \text{if}~\upsilon_r=0~\text{and}~\upsilon_\theta>0 \\
            A_3I_{\sqrt{\upsilon_\theta}}\left(\sqrt{\upsilon_r}r\right)+B_3K_{\sqrt{\upsilon_\theta}}\left(\sqrt{\upsilon_r}r\right) & \text{if}~\upsilon_r>0~\text{and}~\upsilon_\theta>0 \\
           \end{array} \right., 
\end{eqnarray}  
where $\displaystyle\left\{A_3,B_3\right\}$ are arbitrary constants. 

\section{Stress-Displacement Relations}\label{StrssDisRels}

For reference, we write here the linear elastic stress-displacement relations (in cylindrical coordinates), which provide the general mathematical connection between the components of the displacement and stress fields \cite{VTCS2010}:
\begin{subequations}\label{strssdispCYL}
\begin{equation}\label{stssstrnCYL1}
\sigma_{rr}=(\lambda+2\mu){\partial u_r\over\partial r}+{\lambda\over r}\left({\partial u_\theta\over\partial \theta}+u_r\right)+\lambda{\partial u_z\over\partial z},
\end{equation}
\begin{equation}\label{stssstrnCYL2}
\sigma_{\theta\theta}=\lambda{\partial u_r\over\partial r}+{(\lambda+2\mu)\over r}\left({\partial u_\theta\over\partial \theta}+u_r\right)+\lambda{\partial u_z\over\partial z},
\end{equation}
\begin{equation}\label{stssstrnCYL3}
\sigma_{zz}=\lambda{\partial u_r\over\partial r}+{\lambda\over r}\left({\partial u_\theta\over\partial \theta}+u_r\right)+(\lambda+2\mu){\partial u_z\over\partial z},
\end{equation}
\begin{equation}\label{stssstrnCYL4}
\sigma_{r \theta}=\mu\left({1\over r}{\partial u_r\over\partial \theta}+{\partial u_\theta\over\partial r}-{u_\theta\over r}\right)=\sigma_{\theta r},
\end{equation}
\begin{equation}\label{stssstrnCYL5}
\sigma_{rz}=\mu\left({\partial u_r\over\partial z}+{\partial u_z\over\partial r}\right)=\sigma_{zr},
\end{equation}
\begin{equation}\label{stssstrnCYL6}
\sigma_{\theta z}=\mu\left({\partial u_\theta\over\partial z}+{1\over r}{\partial u_z\over\partial \theta}\right)=\sigma_{z\theta}.
\end{equation}
\end{subequations}

\end{document}